\documentclass{aa}
\usepackage{geometry}

\usepackage{caption}
\usepackage{natbib}
\usepackage{amssymb}
\usepackage{enumerate}
\usepackage{graphicx}
\usepackage{rotating}
\usepackage{lscape}

\usepackage{txfonts}
\usepackage{natbib,twoopt}
\usepackage[breaklinks=true]{hyperref} 
\bibpunct{(}{)}{;}{a}{}{,} 
\makeatletter
\newcommandtwoopt{\citeads}[3][][]{\href{http://adsabs.harvard.edu/abs/#3}%
{\def\hyper@linkstart##1##2{}%
\let\hyper@linkend\@empty\citealp[#1][#2]{#3}}}
\newcommandtwoopt{\citepads}[3][][]{\href{http://adsabs.harvard.edu/abs/#3}%
{\def\hyper@linkstart##1##2{}%
\let\hyper@linkend\@empty\citep[#1][#2]{#3}}}
\newcommandtwoopt{\citetads}[3][][]{\href{http://adsabs.harvard.edu/abs/#3}%
{\def\hyper@linkstart##1##2{}%
\let\hyper@linkend\@empty\citet[#1][#2]{#3}}}
\newcommandtwoopt{\citeyearads}[3][][]%
{\href{http://adsabs.harvard.edu/abs/#3}
{\def\hyper@linkstart##1##2{}%
\let\hyper@linkend\@empty\citeyear[#1][#2]{#3}}}
\makeatother

\begin{document}

\title{Identification of dusty massive stars in star-forming dwarf irregular galaxies in the Local Group with mid-IR photometry.
\thanks{Based on observations made with ESO Telescopes at the La Silla Paranal Observatory under programme IDs 090.D-0009 and 091.D-0010.}}

\author{N. E. Britavskiy \inst{1,2} \and A. Z. Bonanos \inst{1} \and  A. Mehner \inst{3} \and M. L. Boyer \inst{4} \and K. B. W. McQuinn \inst{5}}

\offprints{N. Britavskiy}

\institute{IAASARS, National Observatory of Athens, GR-15236 Penteli, Greece\\
 \email{britavskiy@astro.noa.gr, bonanos@astro.noa.gr}
 \and
Section of Astrophysics, Astronomy \& Mechanics, Department of Physics, University of Athens, GR-15783, Athens, Greece
\and
ESO -- European Organisation for Astronomical Research in the Southern Hemisphere, Santiago de Chile, Chile
\and
Observational Cosmology Lab, Code 665, NASA Goddard Space Flight Center, Greenbelt, MD 20771, USA
\and
Minnesota Institute for Astrophysics, School of Physics and Astronomy, University of Minnesota, Minneapolis, MN 55455, USA\\
}
\date{Received 23 April 2015 / Accepted 5 October 2015}

\authorrunning{Britavskiy et al.}
\titlerunning{Identification of dusty massive stars in nearby galaxies}

\abstract
{Increasing the statistics of spectroscopically confirmed evolved massive stars in the Local Group enables the investigation of the mass loss phenomena that occur in these stars in the late stages of their evolution.}
{We aim to complete the census of luminous mid-IR sources in star-forming dwarf irregular (dIrr) galaxies of the Local Group. To achieve this we employed mid-IR photometric selection criteria to identify evolved massive stars, such as red supergiants (RSGs) and luminous blue variables (LBVs), by using the fact that these types of stars have infrared excess due to dust.}
{The method is based on 3.6 $\mu$m and 4.5 $\mu$m photometry from archival {\it Spitzer} Space Telescope images of nearby galaxies. We applied our criteria to 4 dIrr galaxies: Pegasus, Phoenix, Sextans A, and WLM, selecting 79 point sources, which we observed with the VLT/FORS2 spectrograph in multi-object spectroscopy mode.}
{We identified 13 RSGs, of which 6 are new discoveries, also 2 new emission line stars, and 1 candidate yellow supergiant. Among the other observed objects we identified carbon stars, foreground giants, and background objects, such as a quasar and an early-type galaxy that contaminate our survey. We use the results of our spectroscopic survey to revise the mid-IR and optical selection criteria for identifying RSGs from photometric measurements. The optical selection criteria are more efficient in separating extragalactic RSGs from foreground giants than mid-IR selection criteria, however the mid-IR selection criteria are useful for identifying dusty stars in the Local Group. This work serves as a basis for further investigation of the newly discovered dusty massive stars and their host galaxies.}
{}
%
%

\maketitle

\section{Introduction}

Local Group dwarf irregular galaxies with high star-formation rates serve as ideal laboratories for observations of all types of massive stars, as they are typically located in one compact region on the sky and are convenient for observations in multi-object spectroscopy mode. Different properties of dIrrs in the Local Group provide the opportunity to investigate the population and evolution of massive stars within the context of the different metallicities of their host galaxies. The importance of investigating massive stars, especially dusty massive stars, comes from the observed phenomenon of episodic mass loss, which complicates modeling the evolution of these objects \citep{Smith2014}. A census of these stars in nearby galaxies is vital for our understanding of the mass-loss mechanisms and therefore, the different evolutionary stages of massive stars. Thus, increasing the statistics of spectroscopically confirmed dusty massive stars, e.g. RSGs and LBVs, is a prerequisite for studying them further.

The first obstacle in the identification of such rare stars is found in the selection process of these types of objects from photometric catalogs. A systematic search for massive stars in the Local Group became possible with the availability of high quality $BVRI$ photometry. \cite{Massey06,Massey_Images} presented optical photometry for seven star-forming dIrrs, M31 and M33, which served as a basis for further works devoted to the identification of new RSGs \citep{M09,LM2012,Drout12,Levesque13}, LBVs \citep{Massey_Ha,Humphreys2014,Kraus14}, and Wolf-Rayet stars \citep[WR,][]{Neugent2011} in these galaxies. In all these listed works the selection of targets was based on optical $BVRI$ photometry, however, deep optical surveys exist for a small number of dIrrs. Moreover, dusty stars appear brighter in the infrared rather than in optical colors as they exhibit infrared excess. Thus, infrared surveys should, in principle, provide advantages to identifying these types of stars.

We have initiated a survey that aims to provide a census of stars that have undergone episodic mass loss by using available {\it Spitzer} mid-IR photometry to select luminous mid-IR sources in a number of nearby galaxies \citep{Khan10,Khan2011,Khan2013,britavskiy14}. Khan et al. aimed to identify supernova progenitors and objects similar to $\eta~Car$ among the brightest targets in mid-IR colors in a number of nearby galaxies.
The first attempt to systematize the selection criteria for identifying dusty massive stars in nearby galaxies in the mid-infrared was described in \cite{britavskiy14}. As a result of this work 5 new RSGs from 8 selected candidates in the dIrr galaxies Sextans A and IC 1613 were spectroscopically confirmed, therefore providing some evidence for the success of mid-IR selection criteria for dusty massive stars.

In this paper we continue our work on identifying dusty massive stars in the dIrr galaxies of the Local Group. We base our selection approach for interpreting luminous, massive, resolved stellar populations in nearby galaxies at infrared wavelengths on the ``roadmap'' presented by \cite{BMS09,BLK10}. RSGs, supergiant B[e] (sgB[e]) stars, and LBVs are among the brightest infrared sources, due to their intrinsic brightness and due to dust, and they occupy distinct regions in all CMDs.
The paper is organized as follows: in Section 2 we briefly review the procedure for selecting sources, and describe the observations and data reduction of the obtained spectra. In Section 3, we discuss the spectral classification process. Section 4 presents our results, grouped by the object type, in subsections. In Section 5, we discuss our results and evaluate the selection criteria. Section 6 closes the paper with a summary.
\section{Target selection and observations}

\subsection{Target selection}

Following \cite{britavskiy14}, we selected evolved massive star candidates from published {\it Spitzer/IRAC} photometry \citep{Boyer9} of 4 nearby irregular dwarf galaxies with relatively high star formation rates ($\ga 0.003~M_\sun~yr^{-1}$), namely: Pegasus, Phoenix, Sextans A, and WLM.
The selection of galaxies was made by taking into account the occurrence of recent star-formation episodes \citep{Mateo}.
All program galaxies have a few recent star formation populations \citep{Tolstoy1999,Hodge1999}, apart from the low-mass Phoenix galaxy, which has one central, young (100 Myr) stellar population zone \citep{Phoenix_SF}.

LBVs, sgB[e]s, and RSG stars are among the brightest stars at 3.6 $\mu m$ and are concentrated on the CMDs at M$_{3.6 \mu m} < -9$ mag. OB stars are located along a vertical line at $[3.6]-[4.5] \sim 0$ mag. RSGs also fall on this line, but are separated by their extreme brightness, therefore RSG candidates were selected as objects with colors $[3.6]-[4.5] < 0$ and M$_{3.6 \mu m} < -9$ mag. LBVs, sgB[e]s, and WR stars have redder colors than OB stars. In order to find new LBVs and sgB[e] we selected all sources in  \cite{Boyer9} with M$_{3.6 \mu m} \leq -8$ mag and $[3.6]-[4.5] > 0.15$ mag. Additional bright IR sources at smaller $[3.6]-[4.5]$ color  were observed when it was possible to add slits on the FORS2 spectroscopic masks (multi-object MXU mode) without compromising the main targets. Following \cite{BMS09,BLK10}, the selection criteria presented above include LBVs, sgB[e]s, WR stars, and RSGs, but exclude normal OB stars. In the case of the Phoenix galaxy, since all targets were relatively faint, we selected candidates with absolute magnitude M$_{3.6 \mu m} < -6$ mag, in order to fill the rest of the free space on the multi-object mask (MXU).

The selected candidates were cross-correlated with existing optical and near-infrared catalogues to eliminate known foreground stars, H II regions, galaxies, and radio sources. In the end, 43 high priority targets were selected, however, to fill the free space on the MXU masks we added 36 more targets. These additional targets were selected among the brightest sources in the [3.6] band, however, quite often they were fainter than the magnitude cut-off criteria (M$_{3.6 \mu m} \leq -9$ mag), and they did not always satisfy our selection criteria in terms of the $[3.6]-[4.5]$ color. Also, 10 of these additional targets have been reported as candidate carbon stars based on their photometry \citep{Battinelli2000,Battinelli2004,phoenix_carbon}. In total 79 stars were observed in 4 dIrr galaxies: 19 in Pegasus, 14 in Phoenix, 15 in Sextans A, and 31 in WLM.

It should be noted that after the observations were carried out, new mid-IR photometry became available from the DUSTiNGS survey \citep[DUST in Nearby Galaxies with {\it Spitzer}][]{dustings}. This survey includes 3.6 and 4.5-$\mu$m imaging of 50 nearby dwarf galaxies within 1.5 Mpc, aiming to identify dust producing asymptotic giant branch (AGB) stars. Deep images were obtained with the IRAC camera onboard {\it Spitzer} telescope during the post-cryogen phase on 2 epochs, with an average difference of 180 days. Since the new photometry from DUSTiNGS has better accuracy, we used the new [3.6] and [4.5] values (Epoch 1) for our analysis, which vary from the old ones by 0.05 - 0.2 mag in both $[3.6]$ and $[4.5]$ bands, even though the selection process was based on the published photometry from \cite{Boyer9}. Given that the spatial resolution of {\it Spitzer/IRAC} photometry at $[3.6]$ and $[4.5]$ bands is 2", which corresponds to $\approx$ 10 pc at a distance 1 Mpc, it is unlikely to have 2 bright mid-IR sources in this area.

\subsection{Observations and data reduction}

The sample of selected targets was observed with the FORS2 spectrograph at ESO's Very Large Telescope during 2 observing runs (program IDs 090.D-0009 and 091.D-0010). The observations were carried out in multi-object spectroscopy mode, which maximizes the number of selected targets in one observational set. For each galaxy the observations were performed in one MXU field, apart from observations of Sextans A and WLM, where observations were performed in two MXU fields. In Sextans A, one MXU field was observed twice, while for WLM the observations were performed in two different observational fields. The journal of observations and properties of observed galaxies, i.e. the time of observation (MJD), exposure time, distance modulus, systematic radial velocities and metallicities are given in Table \ref{tab:observations}.

The data reduction of the observed targets was performed with the FORS pipeline recipes version 4.9.23 under ESO {\it Reflex} workflows version 2.6 \citep{reflex}. With the help of this pipeline 1D spectra were automatically extracted from the raw image. The reduction process included standard procedures such as bias subtraction, flat field division, background subtraction and wavelength calibration. For each field, observation spectra were obtained four times and combined using the IRAF\footnote{IRAF is distributed by the National Optical Astronomy Observatory, which is operated by the Association of Universities for Research in Astronomy (AURA) under cooperative agreement with the National Science Foundation.} {\it scombine} routine. The final average extracted 1D science spectra were normalized using the {\it continuum} procedure. All spectra were flux calibrated using MOS (moveable slit) mode observations of the flux standard stars. The spectra have an average signal-to-noise ratio (S/N) $\approx$ 20 and a spectral range from 4300 \AA~to 9000 \AA. It should be noted that we did not achieve this wavelength range for all spectra; some of them have a shorter range, which made spectroscopic analysis difficult. The resolving power varies from R $\approx$ 400 at 5000 $\AA$ to R $\approx$ 680 at 8600 $\AA$.

\begin{table*}
\caption{Journal of observations and target galaxy properties.}
\label{tab:observations}
\begin{tabular}{@{}lccccc@{}}
\hline\hline
ID    & MJD &      Exp.        &     Distance modulus  & Radial velocity & $[Fe/H]$\\
      & (JD $-$ 2400000.5)  &        (s)            &  (mag)   &  (km~s$^{-1}$) & (dex) \\
\hline
Pegasus & 56487.34940 & 4x660              & 24.82$\pm$0.07 &  $-183.3$$\pm$5 &   $-1.40$    \\
Phoenix & 56279.13559  & 4x660             & 23.09$\pm$0.10 &  $-13$$\pm$9, $-$52$\pm$6 & $-1.37$  \\
Sextans A (Set 1) & 56360.08635 & 4x660    & 25.78$\pm$0.08 & +324$\pm$2  & $-1.85$\\
Sextans A (Set 2)  & 56423.08622 & 4x660   &  --            & --  & -- \\
WLM (Objects 1-18)       & 56279.08819  & 4x660    & 24.85$\pm$0.08 & $-130$$\pm$1 & $-1.27$ \\
WLM (Objects 19-31)      & 56280.10188 & 4x660     & --  &  --& -- \\
\hline									        	    					
\end{tabular}
\tablefoot{
The distance moduli, systemic radial velocities and metallicities are taken from \cite{all_galax}. The second systematic RV value for Phoenix is from \cite{gallart2012}.}
\end{table*}

\section{Spectral classification process}

The spectral analysis consists of the following steps (i) radial velocity (RV) determination, (ii) spectral type determination and (iii) luminosity class classification. This is the same algorithm used for the identification of RSGs in \cite{britavskiy14}.


We determined the spectral type using the ESO UVES Paranal Observatory Project (POP) library of high-resolution spectra that covers a wide range in luminosity and colors \citep{Bagnulo}. We decreased the resolution of the templates from R $\approx$ 70\,000 to a resolution of R $=$ 440 to match our spectra. This value of the resolution was chosen as an average resolution of the obtained spectra as claimed in the FORS2 spectrograph tutorial.
At this resolution it is quite difficult to distinguish the lines that are indicators of spectral type $-$ instead of lines, we only have blends. Nevertheless, the comparison with templates provides us with the opportunity to investigate the behavior of the TiO bands, which are very strong for K and M spectral types. By using 17 spectral templates with a broad range of spectral types (from F2 to M6), we provide a spectral determination accurate to the level of one late/early spectral subtype.

The determination of luminosity class was based on the investigation of the Ca II line profiles, by comparing the strengths of these gravity-sensitive features with giant and supergiant template spectra. We used the sample of confirmed RSGs and foreground giants in Sextans A from \cite{britavskiy14} as templates. We also used the Ca II triplet for measuring RVs, we cross-correlated spectral templates from the NASA Infrared Telescope Facility spectral library for cool stars \citep{RCV09} against the Ca II region ($\lambda\lambda$ 8380 -- 8800 $\AA$) of our program stars using the IRAF task {\it fxcor}. For cases where the Ca II region was absent in the spectra, the radial velocity analysis was not performed. We used different libraries for the spectral type determination and luminosity class classification because the Ca II triplet region is not included in the ESO UVES POP library.

For a quantitative analysis of the Ca II triplet profiles we used the empirical calibration of the near-IR index CaT \cite[defined by][]{Cenarro01}, which measures the Ca II triplet strength. We calculated indices and errors for each star spectrum where the Ca II triplet existed, by using the indexf package \citep{indexf}. For foreground giants this index (CaT $\backsimeq$ 3 -- 7) usually is half that of supergiants (CaT $>$ 10) and agrees with the empirical fitting-function library \citep{Cenarro02} for stars that have the same spectral type and metallicity. The CaT index measurements confirm the supergiant nature for all stars that have been identified as RSGs from the comparison with template spectra described above.

The resulting spectral type classification of targets  and CaT index, when available, are shown in Tables \ref{tab:phoenix}, \ref{tab:pegasus}, \ref{tab:sexa}, and \ref{tab:wlm}. These tables list a running ID number and ID from the DUSTiNGS of all observed targets, their coordinates, radial velocities (RV), optical magnitude and colors \citep[from][]{Massey_Images}, absolute $[3.6]$ magnitude ($M_{[3.6]}$, computed using the distance in Table \ref{tab:observations}), $[3.6]-[4.5]$ colors, the final spectral classification and notes. In cases where it was not possible to determine the luminosity class, e.g. the spectrum did not include the Ca II triplet region, we provide only the spectral type. In the column "Notes" other spectroscopic information from the literature is listed. The label "Unclassified" was used for the targets with low quality spectra or spectra with high noise that did not allow a spectral type determination.

\section{Results}
\begin{figure*}
\begin{center}
\resizebox{\hsize}{!}{\includegraphics{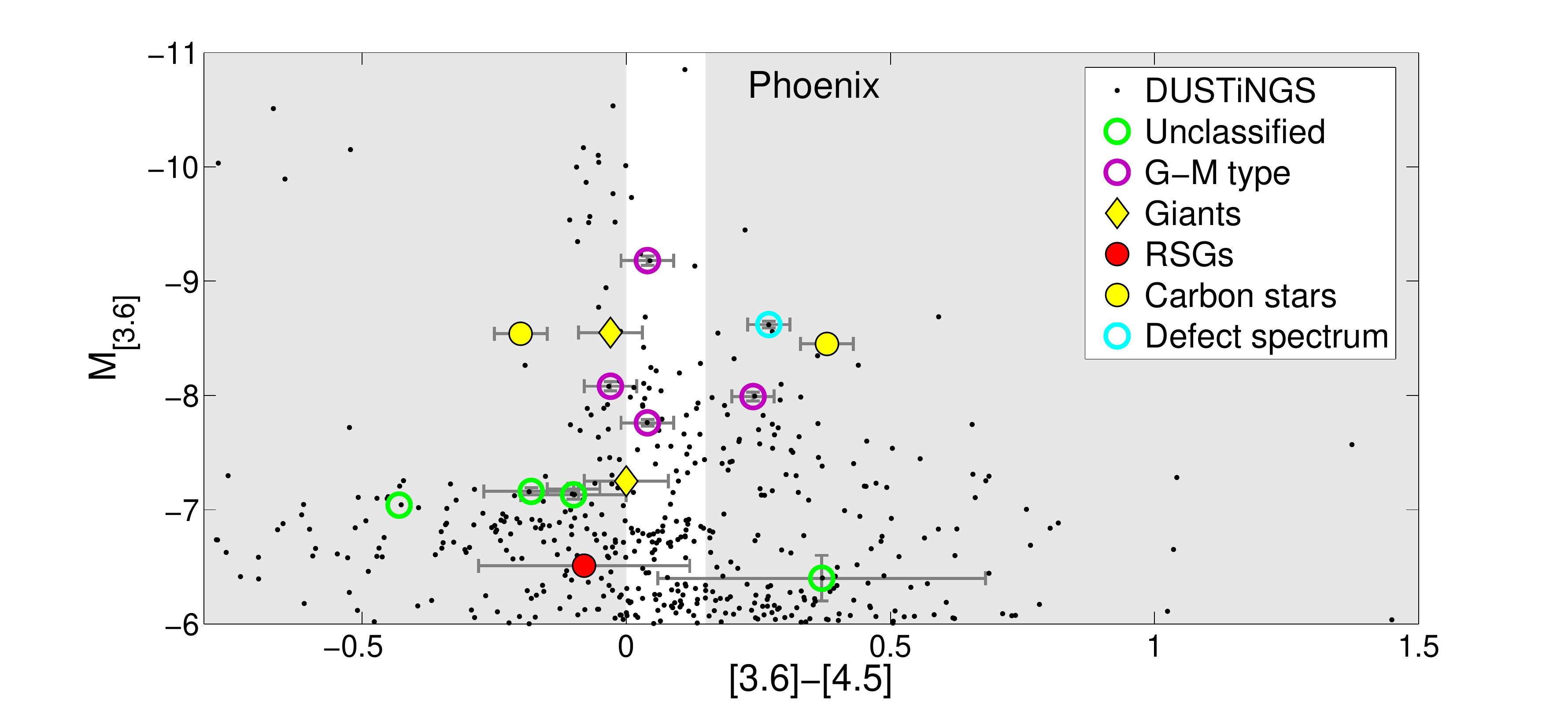}}
\includegraphics[width=0.65\linewidth]{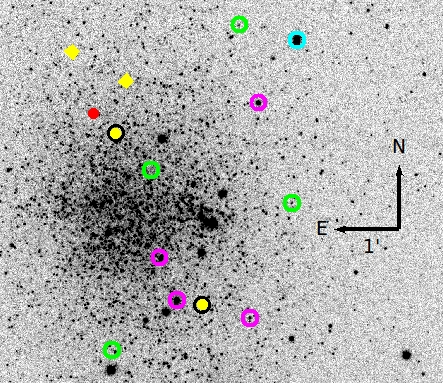}
\end{center}
\caption[]{Top panel: M$_{[3.6]}$ vs. $[3.6]-[4.5]$ CMD for the Phoenix dIrr galaxy. The stars that we have observed are labeled by different symbols according to their classification, given in Table \ref{tab:phoenix} as explained in the legend. For each target on the CMD we indicate the error bars for the colors and magnitudes by grey lines. The regions that satisfy our selection criteria are marked in light grey (the left region corresponds to the RSGs area, the right region corresponds to the LBVs and sgB[e] area, see text for more details). Bottom panel: The spatial distribution of the observed targets, superposed on $V$-band images of the Phoenix galaxy \citep{Massey_Images}.}
\label{Fig1}
\end{figure*}

\begin{figure*}
\begin{center}
\resizebox{\hsize}{!}{\includegraphics{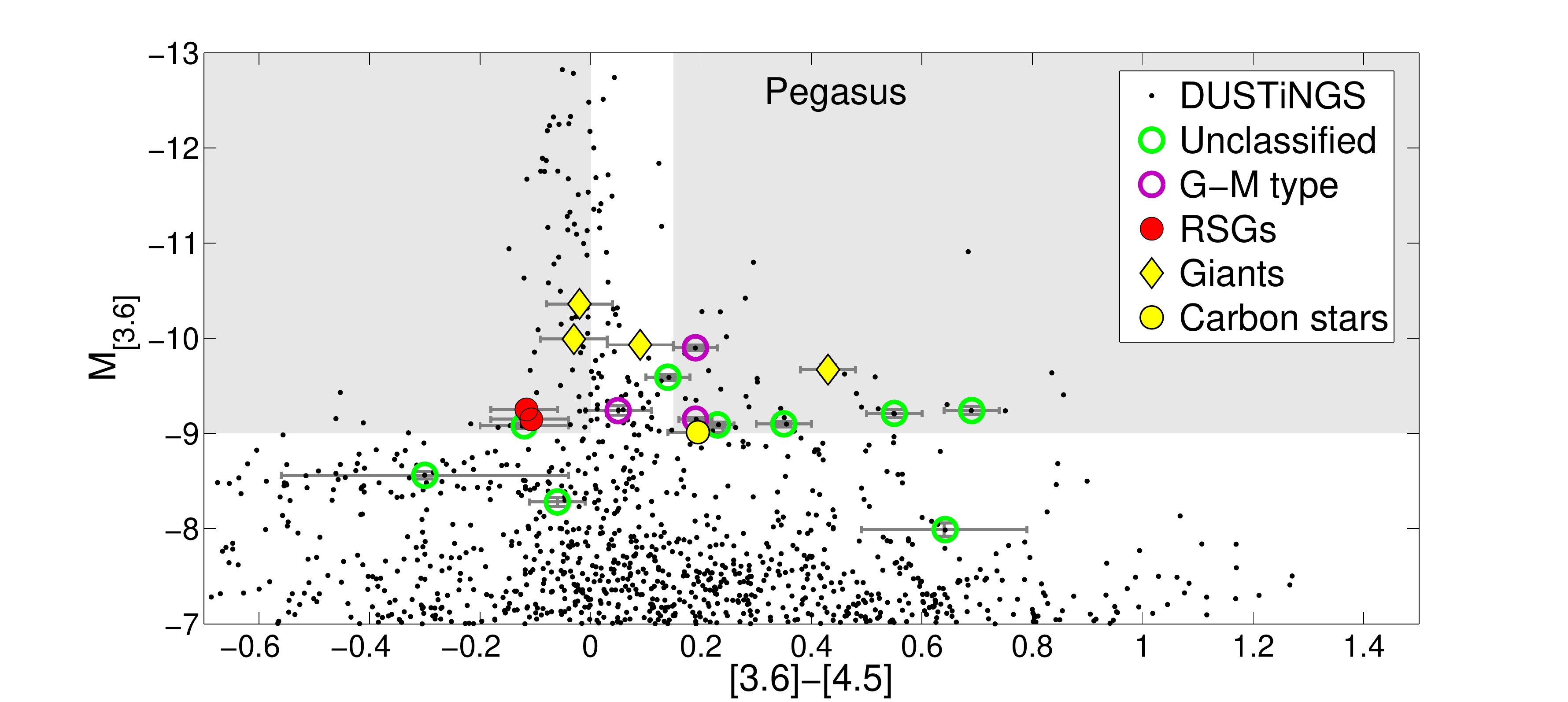}}
\includegraphics[width=0.65\linewidth]{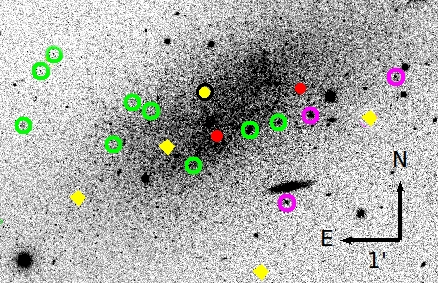}
\end{center}
\caption[]{Same as Figure \ref{Fig1}, but for the Pegasus dIrr galaxy. The stars that we have observed are labeled by different symbols according to their classification, given in Table \ref{tab:pegasus}.}
\label{Fig2}
\end{figure*}

\begin{figure*}
\begin{center}
\resizebox{\hsize}{!}{\includegraphics{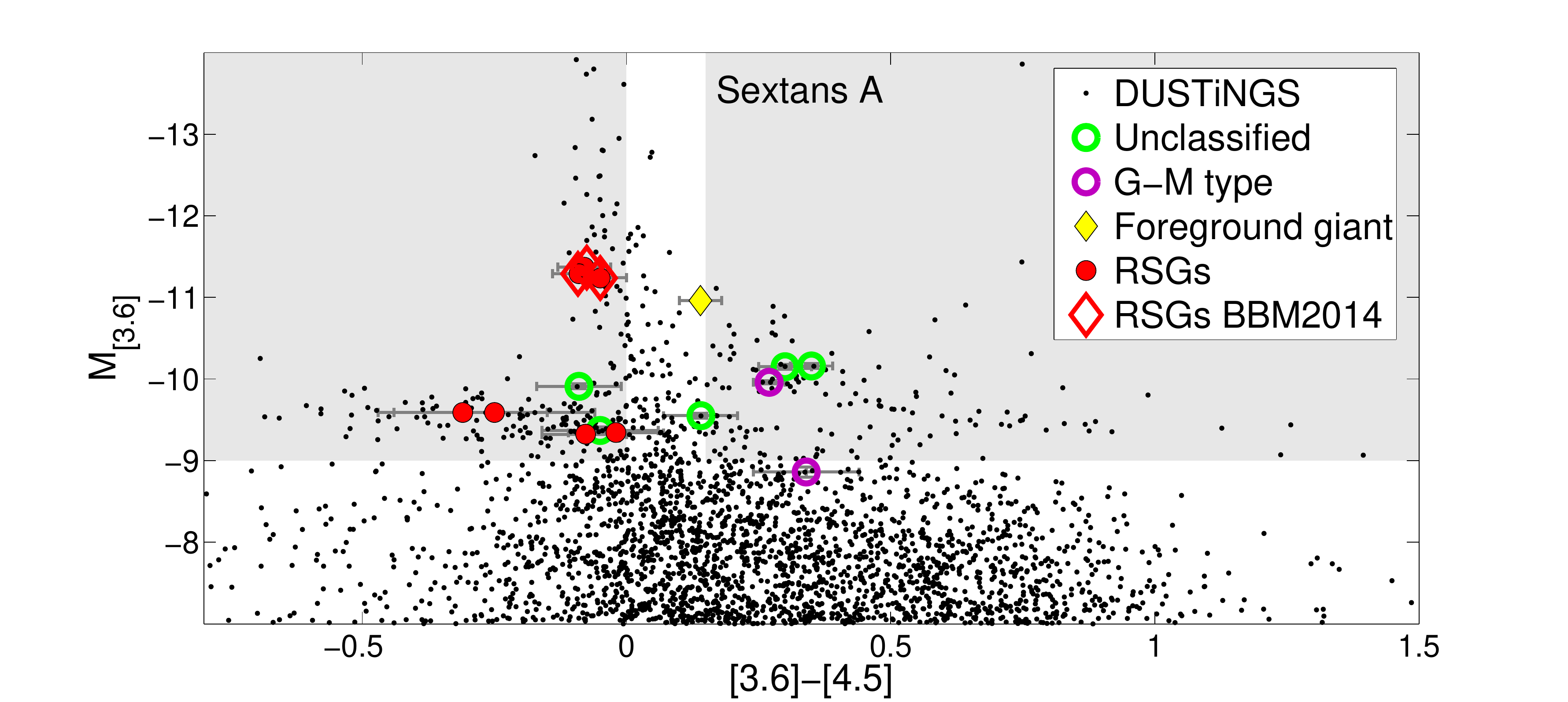}}
\includegraphics[width=0.65\linewidth]{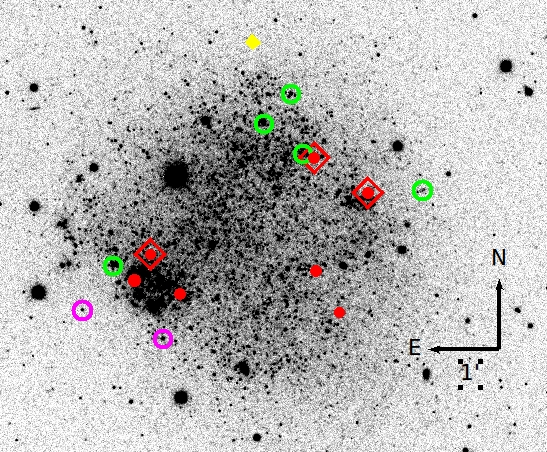}
\end{center}
\caption[]{Same as Figure \ref{Fig1}, but for the Sextans A dIrr galaxy. The stars that we have observed are labeled by different symbols according to their classification, given in Table \ref{tab:sexa}. All previously known RSGs are marked by open red diamonds.}
\label{Fig3}
\end{figure*}

Figures \ref{Fig1} $-$ \ref{Fig4} present the color-magnitude diagrams (CMD) and spatial distribution of our targets in the four program galaxies, Phoenix, Pegasus, Sextans A, WLM, respectively. In the $M_{[3.6]}$ vs. $[3.6] - [4.5] $ CMDs we keep the same formalism of object classification (labels) as we used in Tables \ref{tab:phoenix} $-$ \ref{tab:wlm}. On each CMD we marked the regions that satisfy our photometric selection criteria in light grey. The region on the left part corresponds to the area where RSGs are expected ($[3.6]-[4.5] < 0$), the region on the right corresponds to the LBV and sgB[e] stars ($[3.6]-[4.5] > 0.15$). The value of the absolute magnitude cut-off varies from galaxy to galaxy. For example, the cut-off was decreased to $M_{3.6} < -6$ mag in Phoenix, since all targets in this galaxy are relatively faint in comparison with the other galaxies. In WLM, the cut-off was changed to $M_{3.6} < -8$ mag to select more targets in the LBV region, since 2 MXU masks were available per field. All unclassified objects are faint and have large error bars in the $[3.6]-[4.5]$ color, in comparison with the objects for which spectral classification was performed successfully. In some cases the absolute magnitude of the targets is below the magnitude selection criteria cut-off $M_{3.6} < -9$ mag. This discrepancy is due to the fact that the selection was based on older photometry, while the CMDs plot the more accurate DUSTiNGS photometry.

The analysis of the spatial distribution of observed targets shows us that the majority of foreground and background objects are located outside the main body of dIrrs. Thus, we can conclude that an additional check of the position of candidates with respect to the host galaxy body will help avoid contamination of foreground or background objects.

Among the observed targets we emphasize several groups of dusty massive stars that are described in the subsections below.

\subsection{Red and yellow supergiants}

In Figure \ref{Fig5} we show spectra of all 13 identified RSGs from all galaxies and one candidate yellow supergiant (Sex A 4), ordered by spectral type. Most of the RSGs have K spectral type, apart from 2 M-type RSGs. The low resolution of our spectral data does not allow us to resolve individual lines, which are indicators of luminosity class or spectral type. The main features that are visible is the Ca II triplet and dominant TiO bands in the optical part of RSGs spectra (see Figure \ref{Fig5}). Thus, only the molecular blends are visible in the spectra. In the early-type spectrum of Sex A 4 it was possible to resolve some additional lines. An example of more detailed spectral type and luminosity class identification for red and yellow supergiants from optical wavelength range spectra can be found in \cite{Ginestet94,NMG12,Drout12,britavskiy14}. The spectrum of Sex A 4 has strong hydrogen lines from Paschen series in the Ca II triplet region, however the main feature of this type, the O I 7774 $\AA$ triplet is absent. The presence of strong absorption lines of $H\alpha$ and $H\beta$ in the Sex A 4 spectrum, suggest a G or F spectral type for this target.
In total, we identify 13 RSGs, 6 of which are new discoveries; 4 were previously published in the independent study of \cite{LM2012}, which used optical selection criteria, and 3 RSGs were previously identified by \cite{britavskiy14}. By applying our mid-IR selection criteria we selected these targets independently from \cite{LM2012}. Given two separate spectroscopic observations of these 4 objects, carried out 4 years apart, we can compare their spectral type and check for spectral variability \citep{Massey_var,Levesque10_phys}. Our spectroscopic analysis (see Table \ref{tab:wlm} and Figure \ref{Fig6}) does not show any significant difference in spectral type.

\begin{figure*}
\begin{center}
\resizebox{\hsize}{!}{\includegraphics{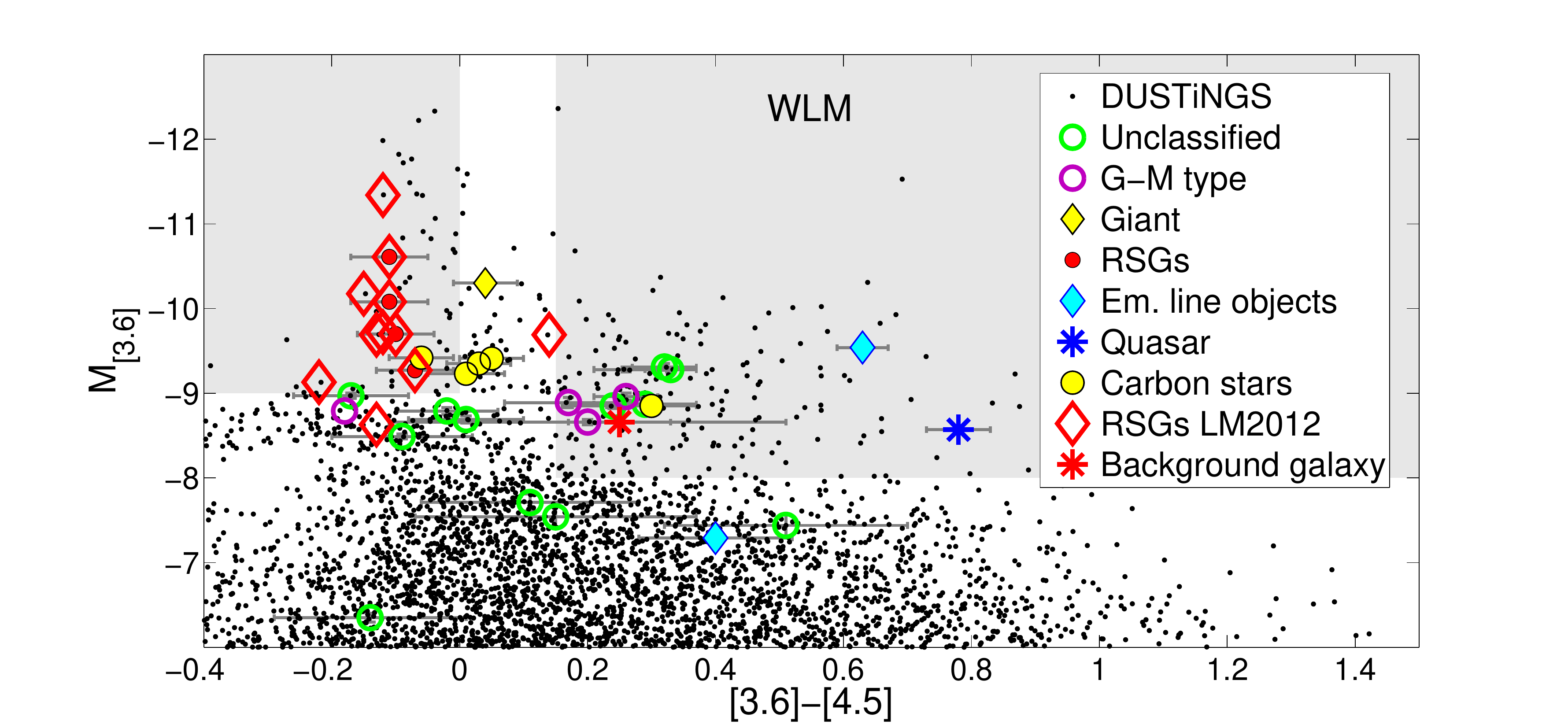}}
\includegraphics[width=0.65\linewidth]{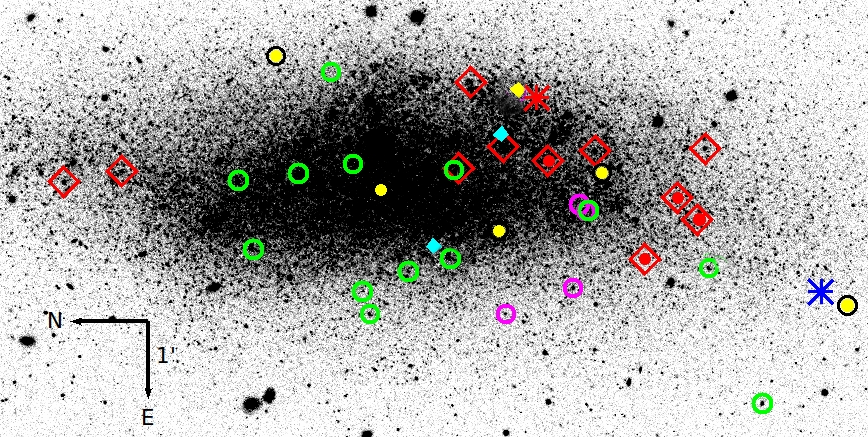}
\end{center}
\caption[]{Same as Figure \ref{Fig1}, but for the WLM dIrr galaxy. The stars that we have observed are labeled by different symbols according to their classification, given in Table \ref{tab:wlm}. Also, all previously known RSGs are marked by open red diamonds, emission line objects are marked by blue filled diamonds, background objects are marked by different color stars.}
\label{Fig4}
\end{figure*}

\begin{figure*}
\resizebox{\hsize}{!}{\includegraphics{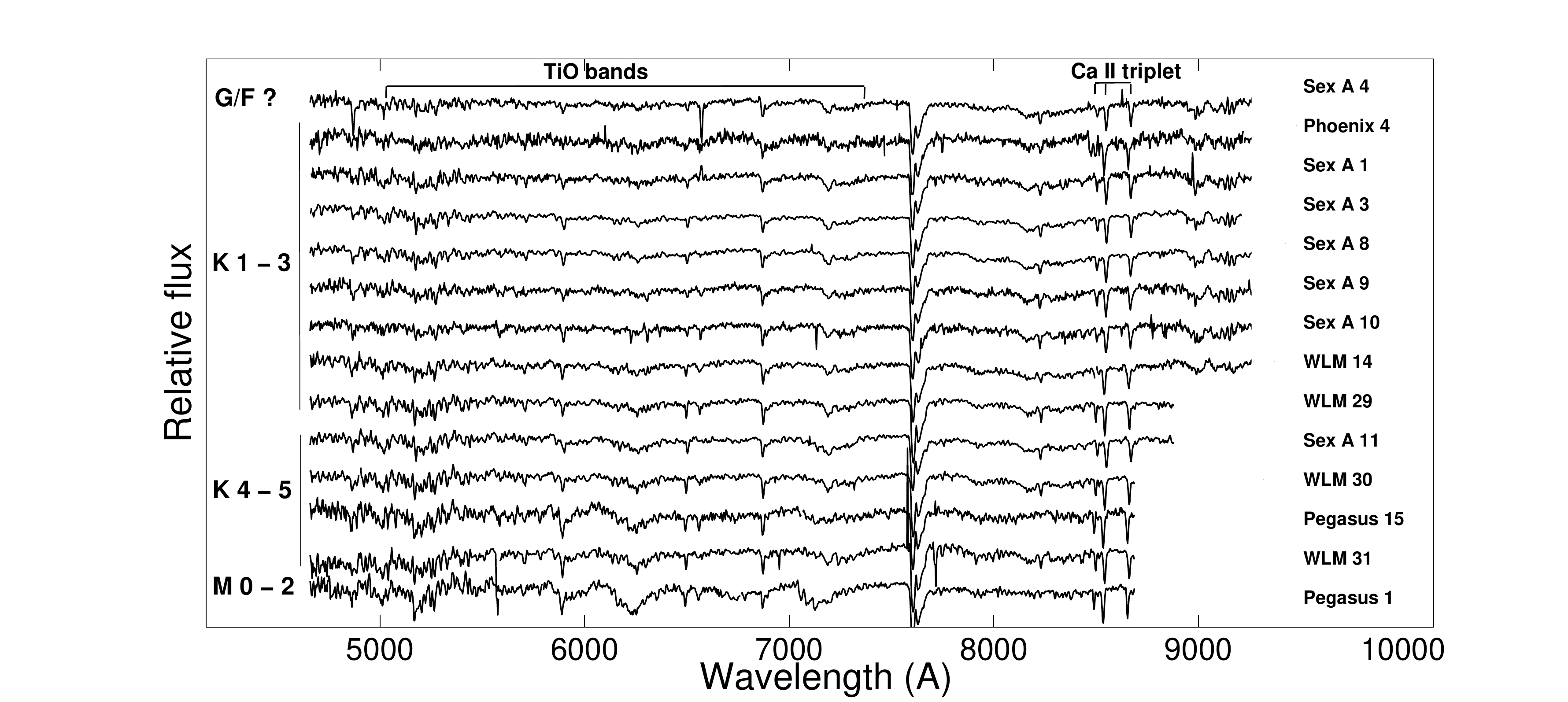}}
\caption[]{Spectra of our 13 RSGs and one candidate yellow supergiant (Sex A 4) in the optical range, ordered by spectral type, labeled on left. Names are given on right. The region where TiO bands dominate, along with the Ca II triplet are marked.}
\label{Fig5}
\end{figure*}

\begin{figure*}
\resizebox{\hsize}{!}{\includegraphics{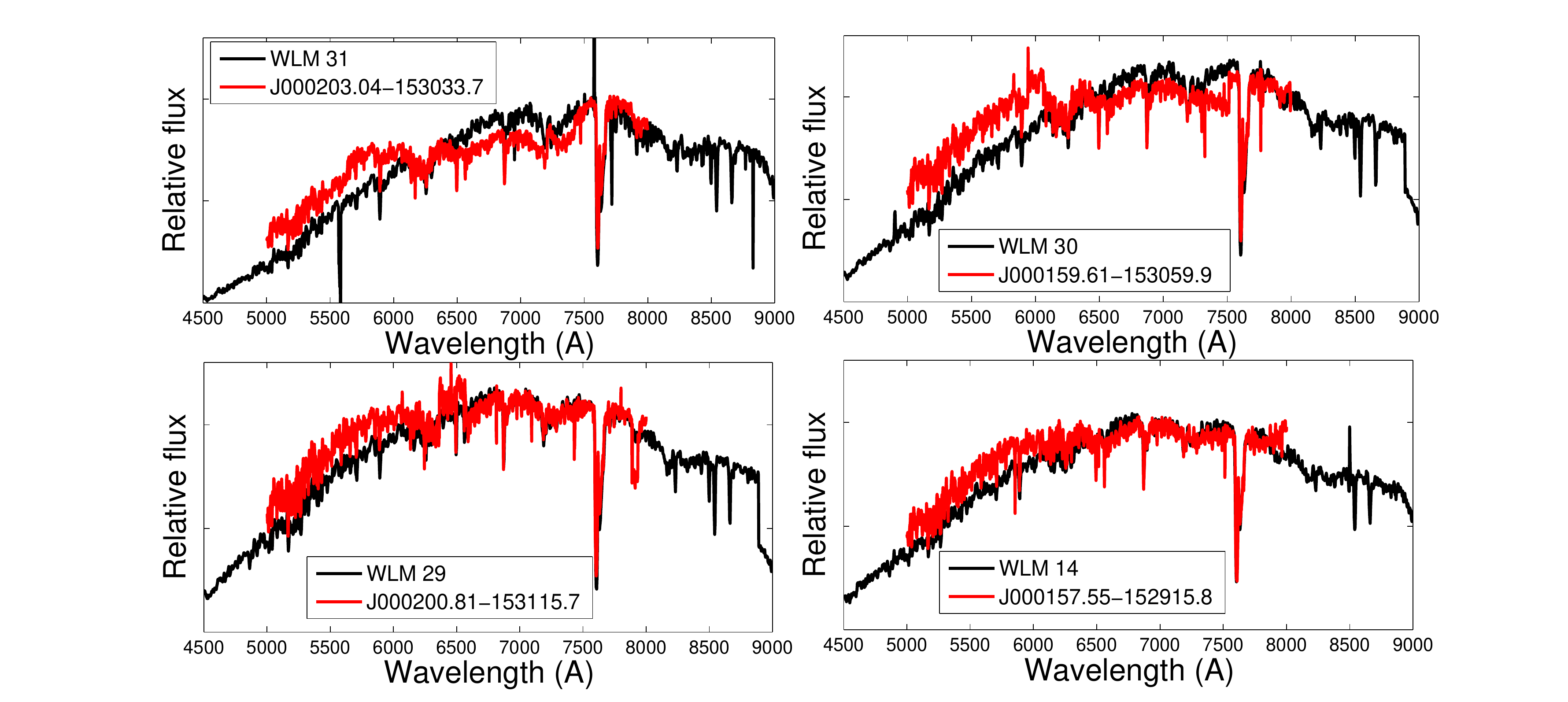}}
\caption[]{Comparison of 4 FORS2 spectra (black) out of 7 previously known RSGs with the spectra of \cite{LM2012} (red).}
\label{Fig6}
\end{figure*}

The positions of all spectroscopically confirmed RSGs on the CMD are predicted by our selection criteria, nevertheless, as we can see in Figure \ref{Fig4} one RSG in WLM \citep[object J000158.14-152332.2 in][]{LM2012} is located on the "red" part ($[3.6] - [4.5] > 0.15$) of the CMD. \cite{britavskiy14} also identified one RSG ("IC 1613 1") with an unusual color, $[3.6] - [4.5] = 0.66$ mag. This color behavior might be explained by a lack of molecular CO and SiO features in the [4.5] band \citep{Verhoelst2009}. Alternatively, the explanation could be found in the photometric variability of these RSGs \citep[see discussion in][]{britavskiy14}.

The low radial velocity values of RSGs in WLM, namely WLM 29, WLM 30, WLM 31, in contrast with other objects from this galaxy (i.e. WLM 14, WLM 17 etc, see Table \ref{tab:wlm}), as well as their spatial distribution (these 3 RSGs are located close to each other, to the south with respect to the main body of WLM, see Fig. \ref{Fig4}) suggest a high rotation velocity for this galaxy. The extragalactic origin of the observed targets was determined by considering the radial velocities, together with the spectral and luminosity class determination, as radial velocities alone do not always distinguish foreground objects from extragalactic ones.

\subsubsection{Foreground giants}

We identified 8 K-M spectral type giants that are foreground stars in the Milky Way. Namely, 2 giants in Phoenix, 4 giants in Pegasus, and 1 giant in both Sextans A and WLM dIrr galaxies. Due to their late spectral types they have the same magnitude and position on the CMD diagrams as RSGs. In the context of the mid-IR selection criteria, giants are hardly distinguishable from RSGs, as discussed further in Section 5. Their classification as giants was based on the relatively weak Ca II triplet and smaller radial velocities compared with RSGs in the same host galaxies.

\subsection{Emission line stars}
\begin{figure*}
\resizebox{\hsize}{!}{\includegraphics{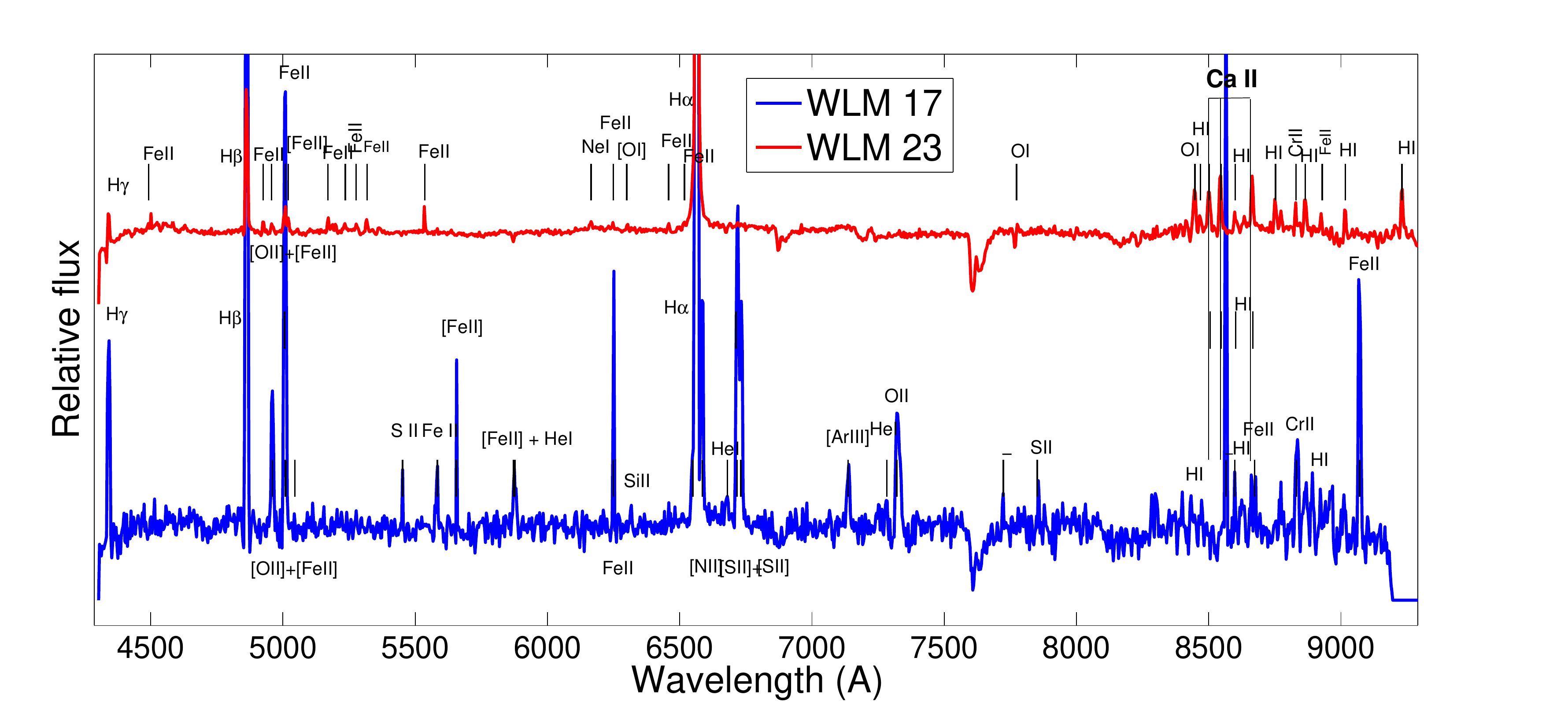}}
\caption[]{Normalized spectra of 2 emission line objects from WLM. Prominent spectral features are labeled.}
\label{Fig7}
\end{figure*}

We discovered 2 rare emission line objects in WLM (WLM 17 and WLM 23), whose spectra are presented in Figure \ref{Fig7}. A list of identified emission lines for each spectrum is presented in Table \ref{tab:eml}. Line identification was performed by using the spectral atlas of $\eta$ Car \citep{eta_linelist} and the spectral atlas of early type hypergiants, including LBVs and sgB[e] stars \citep{chentsov}. The spectral resolution does not allow us to perform an accurate spectroscopic classification of these targets, neither investigate the hydrogen line profiles, however we can provide some information about the nature of these sources.

Besides strong hydrogen lines in emission, object WLM 17 has some forbidden lines of high excitation potential, such as [O II] and [Ar III], which indicate a high temperature of the stars, about $T_{eff}$ $\approx~30\,000$ K. The presence of "nebular" lines [N II] 6583 $\AA$ and [S II] 6717 $\AA$ and 6731 $\AA$, indicates that there is an envelope surrounding this star. The position of this object in the H$\alpha$ image of WLM \citep{Massey_Images}, shows that this object lies inside an HII region. Furthermore, according to \cite{Hodge_HII}, WLM 17 is located near an identified HII region, and spectroscopic features such as [O~III] 5007, He I 5876, N II 6984, He I 6678, [Ar III] 7135 $\AA$, originate from the HII region. We conclude that this object is an evolved, hot, and young massive star near an HII region.

Object WLM 23 has strong hydrogen emission, however the absence of He lines and forbidden Fe lines, indicates a lower temperature for this star, $T_{eff}$ $< 20\,000$ K. The spectrum exhibits both the Ca II triplet and Paschen hydrogen series in emission. Furthermore, dominant Fe II emission lines in the spectrum suggest that this object is an iron star \citep[see Section 2.2 in][]{Humphreys2014}.
These two objects have been selected as candidate LBVs or sgB[e] stars ($[3.6]-[4.6]>0.15$ mag, see Figure \ref{Fig4}). He emission lines and forbidden lines of Fe should exist in these types of stars, but we do not detect them. By taking into account the abundance of Fe II emission lines and the high absolute magnitude ($M_{[3.6]}~= -9.54$ mag), we suggest that object WLM 23 is a hypergiant that belongs to the class of Fe II emission stars, which were defined by \cite{Clark2012}. This object is the first identified Fe II emission line star in the dIrr galaxies of the Local Group.

\subsection{Carbon stars}
\begin{figure*}
\resizebox{\hsize}{!}{\includegraphics{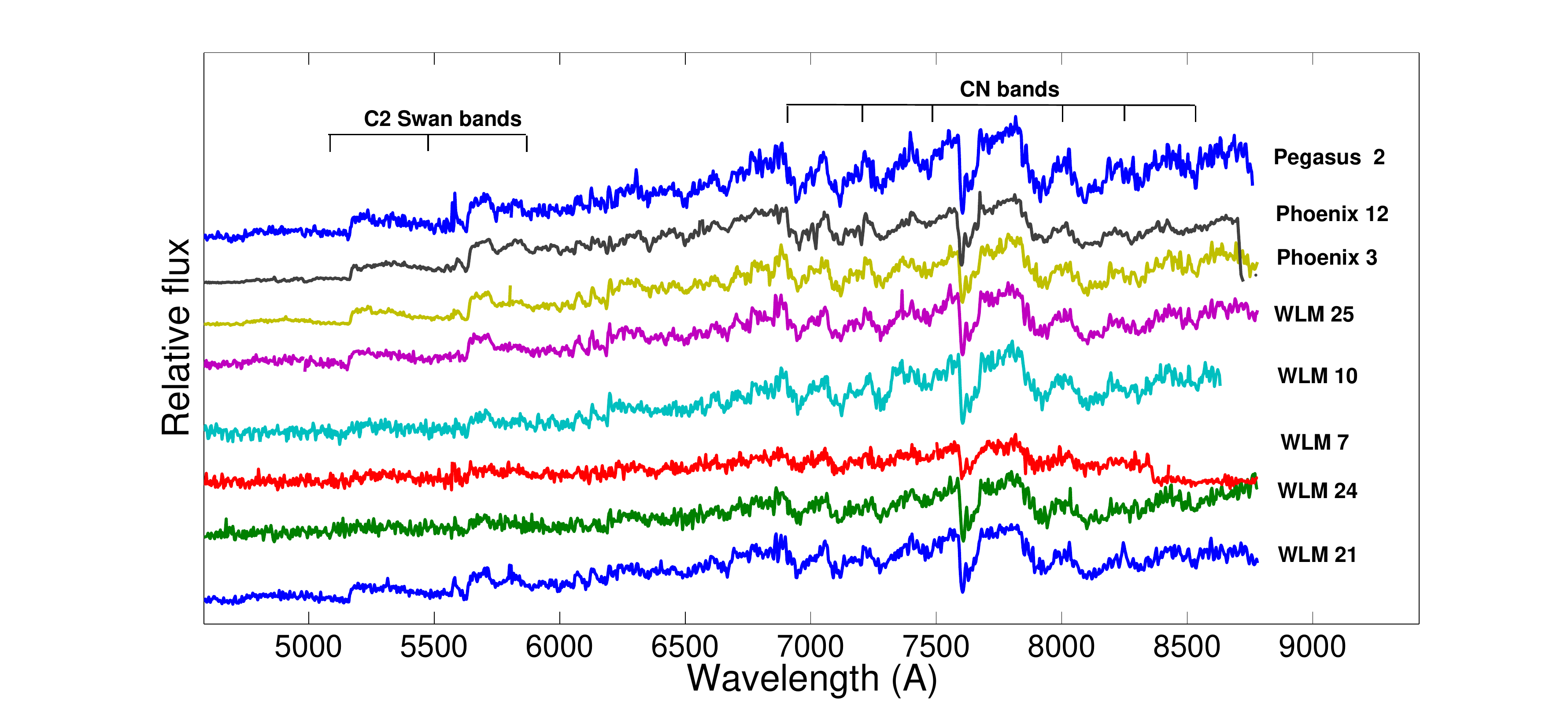}}
\caption[]{Spectra of all carbon stars in our sample, with the main molecular features labeled.}
\label{Fig8}
\end{figure*}

Among the bright sources in mid-IR colors, carbon stars also satisfy our selection criteria. Despite the fact that some candidates were listed in the literature as candidates for carbon stars \citep{Battinelli2000,Battinelli2004,phoenix_carbon}, we still observed them to fill the free space on MXU masks. In total we observed 10 photometric candidates, 8 of which are spectroscopically confirmed as carbon stars in this work and are presented in Figure \ref{Fig8}. For the other 2 carbon star candidates it was not possible to determine the spectral type. The typical spectroscopic molecular features of carbon stars, CN at 6900 -- 7500 $\AA$ and 7900 -- 8400 $\AA$ are detected in all spectra. For a more precise determination of the spectral type we used template spectra of various carbon stars from the SDSS-III archive. From the behavior of the dominant C2 Swan bands in the 4600 $\sim$ 5600 $\AA$ region, we conclude that these objects are all late-type carbon stars, presumably C-N 6-7 series, which, according to the spectral classification in \cite{Keenan_carbon}, are equivalent to early M-type stars. For a diagnostic of the luminosity class we used the ratio of the C2 5165 $\AA$ band to Mg I 5186 $\AA$, which suggests that these objects have a low surface gravity, $log~g$ < 1, corresponding to giants. Some of the identified carbon stars are listed as newly identified variable dusty massive AGB stars, according to \cite{Boyer_var}. These targets are labeled with a comment "VAR" in Tables \ref{tab:phoenix} $-$ \ref{tab:wlm}.


\subsection{Background objects}
\begin{figure*}
\begin{center}
\resizebox{\hsize}{!}{\includegraphics{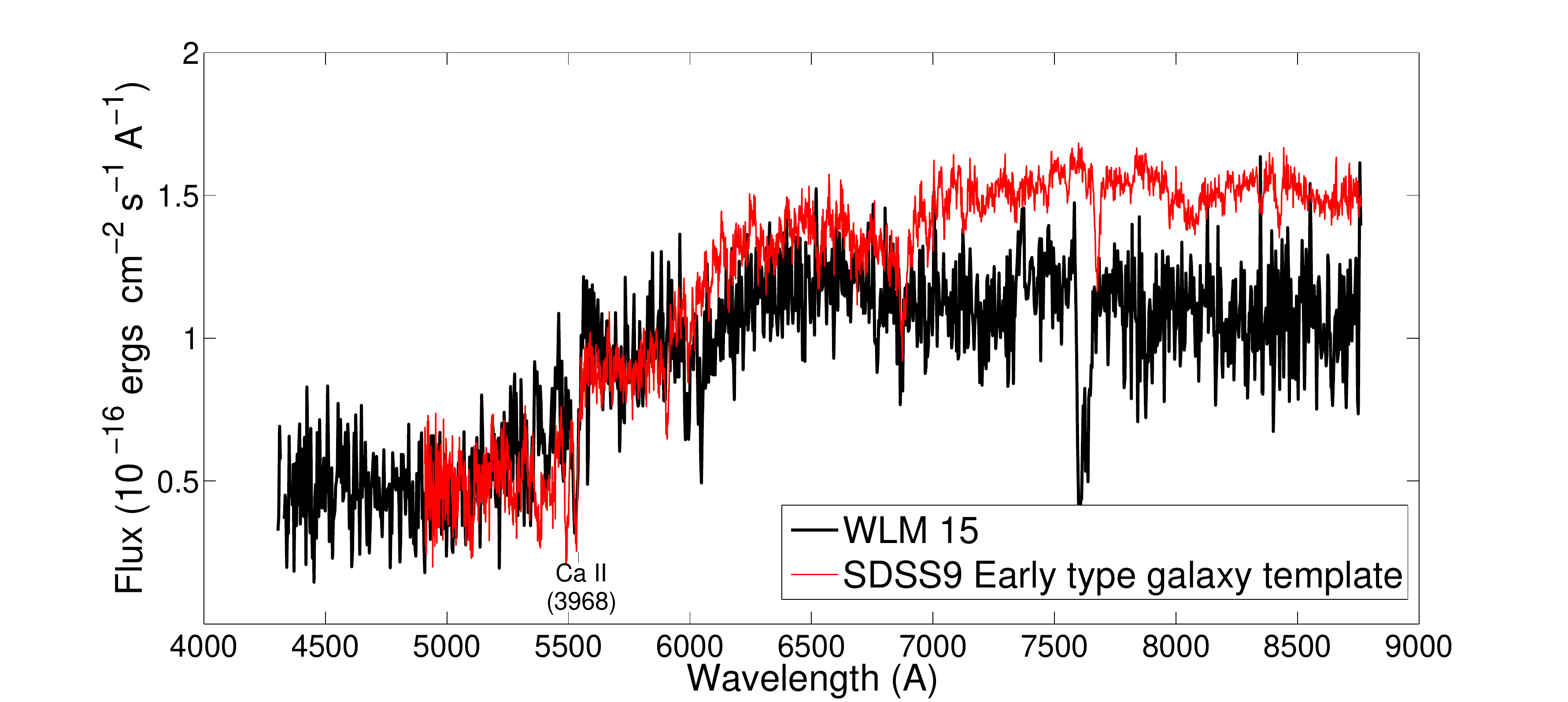}}
\resizebox{\hsize}{!}{\includegraphics{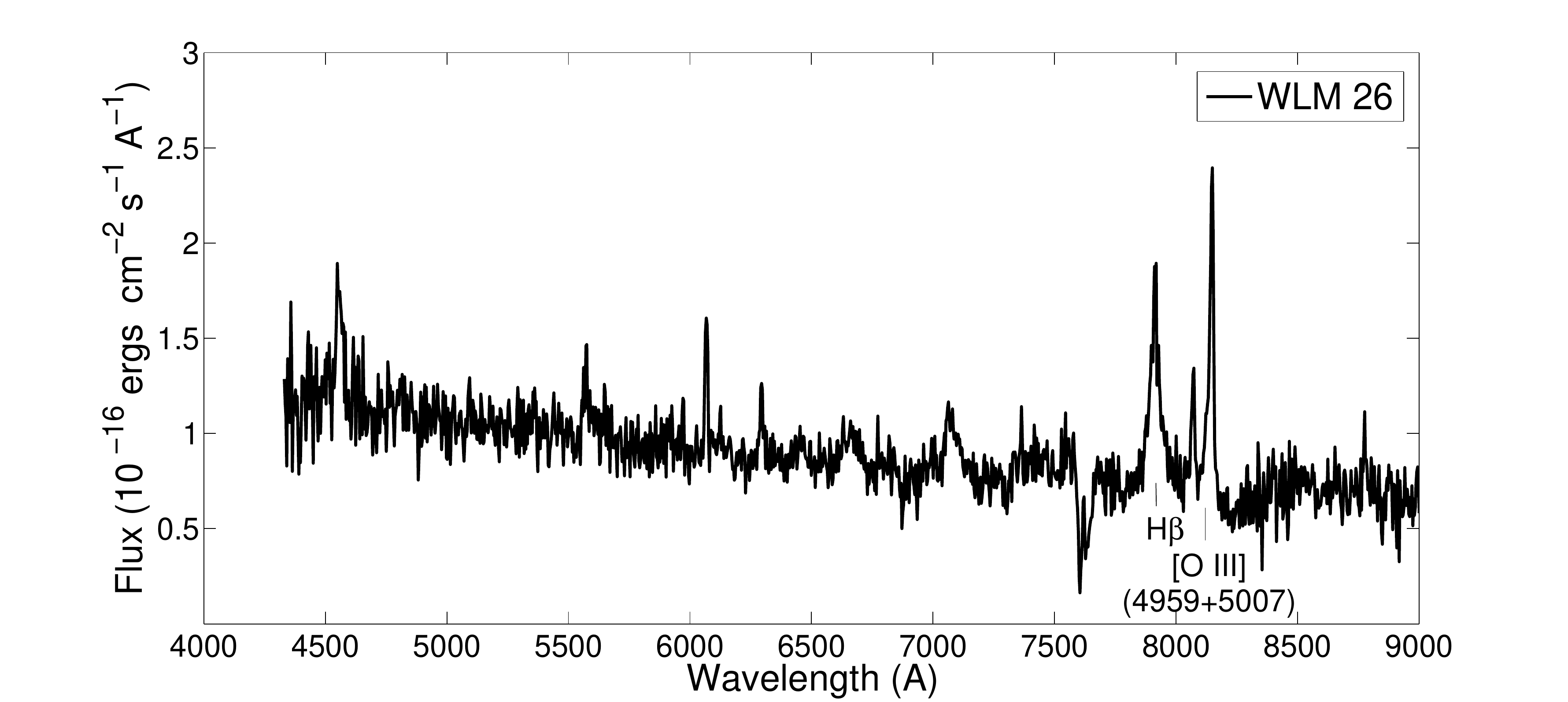}}
\end{center}
\caption[]{Spectra of the background galaxy WLM 15 (upper panel) and quasar WLM 26 (lower panel). An SDSS-III early-type galaxy template (red) is superposed on the spectrum of WLM 15 (black).}
\label{Fig9}
\end{figure*}

We identified two objects among the WLM targets that belong to the background: a quasar (WLM 26) and a background galaxy (WLM 15). By using template spectra from the SDSS-III archive \citep[Data Release 9,][]{sdss_9} we determined the redshift of the quasar to be z = 0.62 based on the H$\beta$ and [O III] lines and the redshift of the galaxy is estimated to be z = 0.39 based on the Ca II 3968 $\AA$ line. The spectra of these two objects are presented in Figure \ref{Fig9}. The absence of emission lines in the spectrum of the galaxy implies that it is an early type galaxy, according to the SDSS-III classification. One of the applications of these objects could be as probes of the interstellar medium in WLM \citep[an example of using radio pulsars for this purpose is presented in][]{Kondratiev2013}. Also, background quasars can be used to determine the proper motion of dIrr galaxies in the Local Group, as done in the Magellanic Clouds \citep{Kozlowski2013}.

\newgeometry{margin=1.5cm}
\begin{landscape}
\begin{table*}
{\small
\caption{Characteristics and classification of observed targets in Phoenix.}
\label{tab:phoenix}
\begin{tabular}{lcccccccrrcr}
\hline\hline
Name &  DUSTiNGS & R.A.(J2000) & Decl.(J2000)  & RV               & $V$	           & $B-V$   & $V-R$   &$M_{[3.6]}$& $[3.6]-[4.5]$ & Spectral                    & Notes 	\\
     & ID   &   (deg)    &  (deg)          &    (km~s$^{-1}$) & (mag)        &  (mag)  & (mag) & (mag)	 &  (mag)	 &	 class  &	  			 \\
\hline
1$^{\dagger}$  &163185 &27.72029&	$-$44.44337&   --	      & $22.27\pm$0.04 & $1.04\pm$0.17 & $0.89\pm$0.05	  &$-7.16\pm$0.03 &$-$0.18$\pm$0.09 & -- &Unclassified	      \\
2$^{\dagger}$  &132464 &27.77338&	$-$44.43438& 	--	     &  --	      &   --	      &       --		  &$-7.04\pm$0.03 &$-$0.43$\pm$0.11& --  &Unclassified       \\
3$^{\dagger}$  &124582 &27.78666&	$-$44.42456&   --	     &    --	      &       --      &       --	  &$-8.54\pm$0.03 &$-$0.20$\pm$0.05  &  Carbon star& MFW2008	     \\
4$^{\dagger}$  &119803 &27.79501&	$-$44.41927&  $-$122$\pm$11& $19.50\pm$0.01 & $1.17\pm$0.01 & $0.64\pm$0.01	  &$-6.51\pm$0.07 &$-$0.08$\pm$0.20 & K1-2 I& CaT=7.8$\pm$0.1  \\
5$^{\dagger}$  &155842 &27.73286&	$-$44.41641&   --	     & $18.04\pm$0.01 & $1.12\pm$0.01 & $0.73\pm$0.01	  &$-8.08\pm$0.04 &$-$0.03$\pm$0.05 & K3-4 &-- 		     \\
6$^{\dagger}$  &126900 &27.78269&	$-$44.41072&   $-$66$\pm$14& $16.44\pm$0.01 & $0.68\pm$0.01 & $0.41\pm$0.01	  &$-8.55\pm$0.04 &$-$0.03$\pm$0.06&  G II, For. &  CaT=3.8$\pm$0.2	    	 \\
7$^{\dagger}$  &115117 &27.80300&	$-$44.40274&   $-$30$\pm$9& $18.74\pm$0.01 & $1.06\pm$0.01 & $0.69\pm$0.01	  &$-7.25\pm$0.04 &$-$0.00$\pm$0.08 & K II, For. &  CaT=3.6$\pm$0.1	  \\
8$^{\dagger}$  &164291 &27.71847&	$-$44.39964& 	--	     & $19.31\pm$0.01 & $1.63\pm$0.02 & $0.57\pm$0.01	  &$-8.62\pm$0.03 & 0.27$\pm$0.04 &--	&      Defect spectrum  	\\
9$^{\dagger}$  &151578 &27.74011&       $-$44.39561& 	--	     &        --      &      --       &        --	  &$-6.40\pm$0.20 & 0.37$\pm$0.31 &	--   & Unclassified         \\
\hline						     		        														
10$^{\dagger}$ &123859 &27.78807&	$-$44.48292&   --	     &        --      &       --      &       --	  &$-7.13\pm$0.04 & $-$0.10$\pm$0.10 &  -- &Unclassified    	\\
11$^{\dagger}$ &153862 &27.73609&	$-$44.47420&   --	     & $20.39\pm$0.01 & $1.41\pm$0.02 & $1.14\pm$0.01	  &$-7.76\pm$0.03 & 0.04$\pm$0.05&	M2-4&--	    	 \\
12             &143442 &27.75407&	$-$44.47075&   --	     &  	--    & 	--    &        --	  &$-8.45\pm$0.03 & 0.38$\pm$0.05&	  Carbon star&MFW2008, VAR 	  \\
13             &138078 &27.76355&	$-$44.46954&   --	     & $19.47\pm$0.01 & $1.49\pm$0.01 & $0.57\pm$0.01	  &$-7.99\pm$0.04 & 0.24$\pm$0.04&   K2-3 &--  	  \\
14             &134407 &27.76993&	$-$44.45805&   --	     & $18.41\pm$0.01 & $1.54\pm$0.01 & $1.05\pm$0.01	  &$-9.18\pm$0.04 & 0.04$\pm$0.05&   M4-6&--  \\
\hline
\end{tabular}
\tablefoot{
$\dagger$ -- targets added to fill the free space on the MXU mask. "MFW2008" refers to photometric candidates for AGB stars, according to \cite{phoenix_carbon}. Targets with the comment "VAR" refer to newly identified dusty variable AGB stars from \cite{Boyer_var}.
For. -- Foreground giants.}
}
\end{table*}
\end{landscape}
\restoregeometry

\newgeometry{margin=1.5cm}
\begin{landscape}
\begin{table*}
{\small
\caption{Characteristics and classification of observed targets in Pegasus.}
\label{tab:pegasus}
\begin{tabular}{lcccccccrrcr}
\hline\hline
Name & DUSTiNGS & R.A.(J2000) & Decl.(J2000)   & RV    & $V$	 & $B-V$   & $V-R$     &  $M_{[3.6]}$& $[3.6]-[4.5]$  &Spectral & Notes	    \\
     & ID &  (deg)    & (deg)              & (km~s$^{-1}$)    & (mag) &  (mag)  & (mag) & (mag)	       &    (mag)      &  class   &	    	\\
\hline
1 &            116706   &352.14938&	14.73709 &$-$257$\pm$18     &$20.72\pm$0.02 & $2.22\pm$0.08  &$1.11\pm$0.02  &$-$9.15$\pm$0.03  &$-$0.11$\pm$0.07 &  M0-2 I	   & CaT=13.4$\pm$0.1	   \\
2 $^{\dagger}$ & 113755 &352.15274&	14.74885 &	--    	      &       --      &      --       &       --       &$-$9.01$\pm$0.04  &  0.19$\pm$0.05  & Carbon star & BD2000	 \\
3 &             111150  &352.15588&	14.72926 &    --      	      &       --      &      --      &        --       &$-$9.08$\pm$0.03  &$-$0.12$\pm$0.08 & --  & Unclassified	 \\
4 &              105161 &352.16320&	14.73414 &$-$121$\pm$18     &$19.55\pm$0.01 &$ 1.22\pm$0.01  &$0.75\pm$0.01  &$-$9.59$\pm$0.05   &	0.43$\pm$0.05 & K3-4 II, For. & CaT=7.3$\pm$0.2 \\
5 $^{\dagger}$  &101247 &352.16760&	14.74385 &	  --  	      &       --      &      --       &       --       &$-$9.21$\pm$0.04 &  0.55$\pm$0.05&  -- & Unclassified, VAR		 \\
6 $^{\dagger}$  & 96974 &352.17285&	14.74603 &	  --  	      &        --      &      --       &      --       &$-$9.10$\pm$0.03 &  0.35$\pm$0.05 & --  & Unclassified	 \\
7 $^{\dagger}$  & 92668 &352.17810&     14.73468 &	   --         &       --      &      --       &        --      &$-$9.24$\pm$0.04 &  0.69$\pm$0.05 & --   & Unclassified		 \\
8 & 84798               &352.18787&	14.72049 &$-$143$\pm$18     &$18.71\pm$0.01 &$ 1.59\pm$0.01  &$1.02\pm$0.01  &$-$10.36$\pm$0.05  &$-$0.02$\pm$0.06& M0-2 II, For. & CaT=5.4$\pm$0.1   \\
9 & 79037               &352.19500&     14.75781 &$-$129$\pm$20     &       --      &      --       &       --       &$-$8.60$\pm$0.03   &	 0.53$\pm$0.05& -- & Unclassified, CaT=4.9$\pm$0.1	    \\
10$^{\dagger}$  & 76368 &352.19830&	14.75453 &	 --	      &$22.51\pm$0.14 &$-0.01\pm$0.19  &$0.30\pm$0.21  &$-$7.99$\pm$0.07   & 0.64$\pm$0.15 & --    & Unclassified	 \\
11$^{\dagger}$ & 72499  &352.20322&	14.73991 &	 --	      &       --      &      --       &        --      &$-$8.56$\pm$0.04   &$-$0.30$\pm$0.26& --  & Unclassified	\\
\hline						  	      		   														
12 &             158441 &352.09976&	14.75298 &	--	      &$20.23\pm$0.01 &$ 1.71\pm$0.03  &$1.08\pm$0.01  &$-$9.24$\pm$0.05 & 0.05$\pm$0.06 &     M0-2  &--		 \\
13 &             152193 &352.10730&	14.74175 &	--    	      &$20.29\pm$0.01 &$ 1.75\pm$0.03  &$1.31\pm$0.01  &$-$9.93$\pm$0.04& 0.09$\pm$0.06 &  M4-6 II, For. & CaT=3.7$\pm$0.1		  \\
14 &             138824 &352.12320&	14.74264 &	 --   	      &$20.12\pm$0.02 &$ 1.79\pm$0.04  &$0.83\pm$0.02  &$-$9.90$\pm$0.03 & 0.19$\pm$0.04 &    Late G & --		   \\
15 &             136294 &352.12616&	14.74971 & $-$248$\pm$25    &$20.68\pm$0.01 &$ 2.07\pm$0.07  &$1.05\pm$0.01  &$-$9.25$\pm$0.04 &$-$0.12$\pm$0.06 &	K4-5 I& CaT=12.1$\pm$0.1 \\
16 &             133060 &352.12994&	14.71921 &	 --	      &$20.16\pm$0.02 &$ 1.49\pm$0.04  &$0.66\pm$0.02  &$-$9.15$\pm$0.02 & 0.19$\pm$0.03 &   K2-3  & --		  \\
17${^\dagger}$ & 131093 &352.13229&	14.74073 &	 --	      &$20.55\pm$0.01 &$ 1.76\pm$0.05  &$0.72\pm$0.02  &$-$8.47$\pm$0.13 & -0.55$\pm$0.25 & -- & Unclassified 	   \\
18 &            127165  &352.13702&	14.70060 &    $-$131$\pm$20 &$18.41\pm$0.01 &$ 1.43\pm$0.01  &$0.87\pm$0.01  &$-$9.99$\pm$0.04 &$-$0.03$\pm$0.06 &  K2-4 III, For. & CaT=6.8$\pm$0.1  \\
19${^\dagger}$ & 124505 &352.14023&	14.73872 &	  --  	      &        --      &       --	&     --	&$-$9.59$\pm$0.03 & 0.14$\pm$0.04 & --  &Unclassified 	     \\
\hline
\end{tabular}
\tablefoot{
$\dagger$ -- targets added to fill the free space on the MXU mask. "BD2000" refers to photometric candidates for AGB stars, according to \cite{Battinelli2000}. Targets with the comment "VAR" refer to newly identified dusty variable AGB stars from \cite{Boyer_var}.
For. -- Foreground giants.}
}
\end{table*}
\end{landscape}
\restoregeometry


\newgeometry{margin=1.5cm}
\begin{landscape}
\begin{table*}
{\small
\caption{Characteristics and classification of observed targets in Sextans A.}
\label{tab:sexa}
\begin{tabular}{lcccccccrrcr}
\hline\hline
Name & DUSTiNGS   & R.A.(J2000) & Decl.(J2000)  & RV		  & $V$  & $B-V$   & $V-R$  &  $M_{[3.6]}$& $[3.6]-[4.5]$  & Spectral& Notes     \\
     & ID      &  (deg)      & (deg)        &(km~s$^{-1}$) & (mag) &  (mag)  & (mag)  &        (mag)  & (mag)   &  class    &		 \\
\hline
 1 & 77330 &	152.76654&	$-$4.70795 & 258$\pm$19     &$20.03\pm$0.01  &$1.76\pm$0.02  &$0.85\pm$0.01 & $-9.59\pm$0.04   &$-$0.31$\pm$0.16 &  K1-3 I  & CaT=12.7$\pm$0.2\\
 2 & 74652 &	152.77032&	$-$4.71814 &   --	      &$20.70\pm$0.01  &$1.62\pm$0.06  &$0.85\pm$0.02 & $-9.96\pm$0.02   & 0.27$\pm$0.03   &  Late G$-$K&  --		            \\
 3 & 72683 &	152.77316&	$-$4.69916 &  291$\pm$19    &$18.32\pm$0.01  &$1.80\pm$0.01  &$0.91\pm$0.01 &$-11.37\pm$0.03  &$-$0.08$\pm$0.05 &   K1-3 I	&Early K RSG, BBM2014, CaT=11.5$\pm$0.1	      \\
 4 & 70373 &	152.77670&	$-$4.70510 & 258$\pm$21     &$19.58\pm$0.01  &$1.44\pm$0.02  &$0.77\pm$0.01 & $-9.59\pm$0.04   &$-$0.25$\pm$0.19 &   G-F? & YSG candidate, CaT=11.4$\pm$0.1		 \\
 5 & 67272 &	152.78140&	$-$4.70179 &	 --   	      &$20.90\pm$0.01  &$1.09\pm$0.06  &$0.69\pm$0.02 & $-9.55\pm$0.03   & 0.14$\pm$0.07   & --  &Unclassified			    \\  		
 6 & 62499 &	152.78842&	$-$4.71166 &	 --   	      & --	       & --		& --		&$-8.86\pm$0.06 &  0.34$\pm$0.10  &	  Late G-K&  --			     \\ 		 
\hline					    	      		 															
7 & 115101 &	152.71196&	$-$4.68491 &	     --       & --		 & --		&     --      &$-9.37\pm$0.04  &$-$0.05$\pm$0.11 & --  & Unclassified				      \\
8 & 106505 &	152.72426&	$-$4.68539 &	 217$\pm$14 &$18.29\pm$0.01  &$1.86\pm$0.01  &$0.93\pm$0.01 &$-11.29\pm$0.03  &$-$0.09$\pm$0.05 &	  K1-3 I	& Early K RSG, BBM2014, CaT=12.4$\pm$0.1   \\
9 & 102187 &	152.73050&	$-$4.71217 &	 200$\pm$22 &$19.93\pm$0.01  &$1.58\pm$0.02  &$0.81\pm$0.01 & $-9.34\pm$0.05   &$-$0.02$\pm$0.09 &	  K1-3 I& CaT=11.8$\pm$0.1	   \\ 		
10 & 98470 &	152.73587&	$-$4.70284 &	 214$\pm$27 &$19.85\pm$0.01  &$1.45\pm$0.01  &$0.75\pm$0.01 & $-9.32\pm$0.04   &$-$0.08$\pm$0.08 &	 K1-3 I& CaT=11.0$\pm$0.3 \\			
11 & 98112 &	152.73636&	$-$4.67753 &	 281$\pm$23 &$18.59\pm$0.01  &$1.89\pm$0.01  &$0.99\pm$0.01 &$-11.24\pm$0.03  &$-$0.05$\pm$0.05 &	 K3-5 I 	& Late K RSG, BBM2014, CaT=12.1$\pm$0.2 \\
12 & 96477 &	152.73883&	$-$4.67685 &	   -- 	      &$19.35\pm$0.01  &$1.52\pm$0.01  &$0.79\pm$0.01 & $-9.91\pm$0.03   &$-$0.09$\pm$0.08 &-- &  Unclassified 				\\
13 & 94601 &	152.74149&	$-$4.66321 &	     --       &$20.00\pm$0.01  &$1.59\pm$0.03  &$0.70\pm$0.01 &$-10.15\pm$0.03   & 0.30$\pm$0.05 & -- & Unclassified				 \\	
14 & 90437 &	152.74760&	$-$4.66993 &	   -- 	      &$20.56\pm$0.02  &$1.34\pm$0.08  &$0.64\pm$0.02 &$-10.16\pm$0.03   & 0.35$\pm$0.04 &  --  &Unclassified				  \\
15 & 88681 &	152.75012&	$-$4.65169 &	  --  	      &$19.09\pm$0.01  &$1.48\pm$0.01  &$0.98\pm$0.01 &$-10.96\pm$0.03  & 0.14$\pm$0.04 &  M0-2 II, For. &  Late M giant, BBM2014			  \\
\hline																	
\end{tabular}\\																
\tablefoot{
"BBM2014" -- refers to targets previously observed by \cite{britavskiy14}. For. -- Foreground giants.}
}
\end{table*}																
\end{landscape}		
\restoregeometry																			

\newgeometry{margin=1.5cm}
\begin{landscape}
\begin{table*}
{\small
\caption{Characteristics and classification of observed targets in WLM.}
\label{tab:wlm}
\begin{tabular}{lcccccccrrcr}
\hline\hline
Name & DUSTiNGS & R.A.(J2000) & Decl.(J2000)   &    RV	            & $V$      & $B-V$   & $V-R$   &  $M_{[3.6]}$& $[3.6]-[4.5]$  & Spectral  & Notes         \\
     & ID       &  (deg)       &  (deg)      &     (km~s$^{-1}$)    & (mag) &  (mag)     & (mag)   & (mag)         &  (mag)	 & class  &		           \\
\hline
  1            & 81399   &	0.51547&  $-$15.45649  &     -- 	 &$22.26\pm$0.04 &$ 0.97\pm$0.09 &$ 1.17\pm$0.05   & $-8.49\pm$0.03& $-$0.09$\pm$0.11 & --  &Unclassified, 2 obs.	\\
  2            & 74034  &	0.52545&  $-$15.44786  &     -- 	 &$21.87\pm$0.03 &$ 1.15\pm$0.11 &$0.48\pm$0.05   & $-8.48\pm$0.04&  $-$0.25$\pm$0.11& --   &Unclassified	       \\
 3${^\dagger}$ & 77883  &	0.52019&  $-$15.44624  &    --  	 &   -- 	 &  --  	 & --		   & $-8.97\pm$0.03& $-$0.17$\pm$0.09& -- &Unclassified        \\
 4${^\dagger}$ & 100786 &	0.49050&  $-$15.44401  &    --  	 &$22.50\pm$0.09 &$ 0.27\pm$0.13 &$0.70\pm$0.09   & $-6.10\pm$0.12& 0.11$\pm$0.16& -- &Unclassified	       \\
 5${^\dagger}$ & 117303 &	0.46916&  $-$15.43916  &     --       	 &   -- 	 &  --  	 & --		   & $-8.87\pm$0.04& 0.29$\pm$0.08&  --   &Unclassified 	 \\
 6${^\dagger}$ & 99211  &	0.49273&  $-$15.43183  &    --  	 &$21.64\pm$0.01 &$ 1.47\pm$0.12 &$ 0.93\pm$0.02   & $-6.83\pm$0.20& 0.61$\pm$0.25 & --  &Unclassified\\
 7${^\dagger}$ &120017 &	0.46540&  $-$15.42686  &    --  	 &   -- 	 &  --  	 & --		   & $-8.85\pm$0.04& 0.30$\pm$0.07 & Carbon star& BD2004, VAR		 \\
 8${^\dagger}$ &85266  &	0.51034&  $-$15.42175  &   --		 &$22.33\pm$0.05 &$ 0.31\pm$0.07 &$ 0.48\pm$0.06   & $-8.69\pm$0.04& 0.01$\pm$0.09 & --  &Unclassified  	\\
 9${^\dagger}$ &97873  &	0.49439&  $-$15.41839  &    --  	 &$21.42\pm$0.02 &$ 0.50\pm$0.03 &$ 0.89\pm$0.02   & $-8.85\pm$0.03& 0.24$\pm$0.08 & --  &Unclassified, VAR		\\
\hline									  															
 10${^\dagger}$ & 99256&	0.49260&  $-$15.49980  &      --	 &$22.61\pm$0.03 &$-0.28\pm$0.04 &$-0.03\pm$0.05   & $-9.23\pm$0.04  & 0.01$\pm$0.06 &   Carbon star &BD2004, 2 obs.	\\
 11${^\dagger}$ &  92333 &	0.50126&  $-$15.49678  &   --	      	 &$21.39\pm$0.02 &$ 1.58\pm$0.06 &$ 0.96\pm$0.02   & $-7.44\pm$0.13  & 0.51$\pm$0.19   &  -- & Unclassified		 \\
 12${^\dagger}$ & 93455 &	0.49988&  $-$15.49480  &   --	      	 &$22.14\pm$0.02 &$-0.29\pm$0.03 &$-0.05\pm$0.04   & $-8.89\pm$0.06  & 0.17$\pm$0.10 & Late G	  & 2 obs., VAR   \\
 13 &            78469  &	0.51934&  $-$15.49346  &   --	      	 &$21.01\pm$0.01 &$ 1.51\pm$0.07 &$ 0.99\pm$0.02   & $-8.96\pm$0.04  & 0.26$\pm$0.05&	   G-K    &	  --		   \\
 14&            101523 &	0.48976&  $-$15.48786  & $-$142$\pm$26   &$19.26\pm$0.01 &$ 1.64\pm$0.01 &$ 0.86\pm$0.01   & $-9.27\pm$0.04& $-$0.07$\pm$0.06&  K1-3 I & K0-1 I, LM2012, 2 obs., CaT=9.9$\pm$0.3    \\
 15${^\dagger}$ & 112858  &	0.47521&  $-$15.48523  & --	      	 &$21.01\pm$0.01 &$ 1.60\pm$0.03 &$ 0.96\pm$0.01   & $-8.66\pm$0.03& 0.25$\pm$0.08 & Back. galaxy & z=0.39, 2 obs.		   \\
 16${^\dagger}$ & 113729  &	0.47403&  $-$15.48050  &     $-$40$\pm$3 &$18.83\pm$0.01 &$ 1.50\pm$0.01 &$ 1.03\pm$0.01   &$-10.30\pm$0.04& 0.04$\pm$0.05 &	    M1-3 II, For. & 2 obs., CaT=5.2$\pm$0.1	   \\	
17 &              106415 &	0.48364&  $-$15.47721  & $-$145$\pm$4	 &$21.31\pm$0.01 &$ 0.10\pm$0.02 &$ 0.22\pm$0.02   & $-7.29\pm$0.08& 0.40$\pm$0.12&   Em. l. star&$H\alpha$ source, MMO2007	    \\
18 &             99700 &	0.49205&  $-$15.46671  &     --       	 &$20.54\pm$0.01 &$ 1.36\pm$0.01 &$ 0.74\pm$0.01   & $-9.28\pm$0.03& 0.33$\pm$0.04&  -- &Unclassified, 2 obs.		    \\
\hline	
 19 &            92197 &	0.50146&  $-$15.49637  &    --        	 &$21.39\pm$0.02 &$ 1.58\pm$0.06 &$ 0.96\pm$0.02   & $-8.79\pm$0.04&	  -0.18$\pm$0.10&	   K& --			       \\
 20 &            74091 &	0.52535&  $-$15.47827  & $-$214$\pm$14   &$21.68\pm$0.01 &$ 1.64\pm$0.15 &$ 1.18\pm$0.02   & $-8.66\pm$0.03&   0.20$\pm$0.31		   & K& --		       \\	
 21${^\dagger}$ & 88592 &	0.50603&  $-$15.47669  &   --	      	 &$21.81\pm$0.02 &$ 1.55\pm$0.11 &$ 1.11\pm$0.03   & $-9.37\pm$0.05&	   0.03$\pm$0.05& Carbon star& BD2004		     \\
 22${^\dagger}$ & 83875 &       0.51208&  $-$15.46514  &   $-$145$\pm$4  &   -- 	 &  --  	 & --		   & $-9.31\pm$0.04&	   0.32$\pm$0.05& Unclassified& BD2004, 2 obs., VAR	\\
 23             & 85741 &       0.50967&  $-$15.46217  &   $-$135$\pm$5  &$19.45\pm$0.01 &$ 0.09\pm$0.01 &$ 0.37\pm$0.01  & $-9.54\pm$0.03&   0.63$\pm$0.04&  Em. l. star &$H\alpha$ source, MMO2007	  \\
 24${^\dagger}$ & 96047 &	0.49661&  $-$15.45020  &  --	      	 &$21.93\pm$0.02 &$-0.38\pm$0.03 &$-0.02\pm$0.03   & $-9.42\pm$0.03&$-$0.06$\pm$0.05 &     Carbon star& BD2004        \\
\hline									  															
25${^\dagger}$ & 75513&  	0.52343&  $-$15.55465  & --	      	 &   -- 	 &  --  	 & --		   & $-9.41\pm$0.03&   0.05$\pm$0.05&	    Carbon star& BD2004        \\
 26 &           77794 &  	0.52028&  $-$15.54894  &    --        	 &$21.45\pm$0.01 &$ 0.56\pm$0.02 &$ 0.45\pm$0.01   & $-8.57\pm$0.04&   0.78$\pm$0.05&	      Quasar& z=0.62	       \\	
 27 &           58933 &	        0.54626&  $-$15.53583  &   --	      	 &$20.17\pm$0.01 &$ 1.40\pm$0.01 &$ 0.67\pm$0.01   & $-8.79\pm$0.04&   $-$0.02$\pm$0.08& --  &Unclassified	       \\	
 28${^\dagger}$ &82225 &	0.51428&  $-$15.52568  & --	      	 &$21.12\pm$0.01 &$ 0.34\pm$0.01 &$ 0.26\pm$0.01   & $-9.28\pm$0.03&	   0.26$\pm$0.05&   Unclassified & BD2004, VAR  	 \\
 29 &            90598 &	0.50340&  $-$15.52166  &   $-$28$\pm$11  &$18.69\pm$0.01 &$ 1.78\pm$0.01 &$ 0.91\pm$0.01   &$-10.08\pm$0.04&$-$0.11$\pm$0.06&	 K1-3 I & K0-1 I, LM2012, CaT=13.1$\pm$0.1		 \\
 30 &            94581 &	0.49837&  $-$15.51678  &   $-$44$\pm$13  &$18.97\pm$0.01 &$ 1.78\pm$0.01 &$ 0.90\pm$0.01   & $-9.70\pm$0.03&$-$0.10$\pm$0.06&	K3-5 I & K2-4 I, LM2012, CaT=12.9$\pm$0.1	 \\
 31 &            83414 &	0.51268&  $-$15.50950  &   $-$48$\pm$16  &$18.67\pm$0.01 &$ 1.97\pm$0.01 &$ 1.04\pm$0.01   &$-10.61\pm$0.04&$-$0.11$\pm$0.06&	   K4-5 I & K5 I, LM2012, CaT=13.8$\pm$0.1		   \\
\hline
\end{tabular}
\tablefoot{
$\dagger$ -- targets added to fill the free space on the MXU mask. Targets with the following comments are: "BD2004" -- \cite{Battinelli2004}, "LM2012" -- \cite{LM2012}, "MMO2007" -- \cite{Massey_Ha}. "2 obs." -- targets observed twice in different masks. Targets with the comment "VAR" refer to newly identified dusty variable AGB stars from \cite{Boyer_var}. For. -- Foreground giants.}
}
\end{table*}
\end{landscape}
\restoregeometry


\begin{table}
\caption{Identified emission lines in WLM 17 and WLM 23}
\label{tab:eml}
\begin{tabular}{lc|lc}
\hline\hline
\multicolumn{2}{c}{WLM 17} &\multicolumn{2}{c}{WLM 23} \\
\hline
Ion     & Wavelength  &   Ion     & Wavelength  \\
         &  (A)       &           &  (A)           \\
\hline
$H\gamma$     &4340                 &$H\gamma$ &4340    \\
$H\beta $   &4861                   &   --	 &4501         \\
$[O~II]+[Fe~II]$ or  & 4957   &$H\beta$  &4861      \\
$[O~III]$ ?     &    4960         &$Fe~II$      &4925     \\
$[Fe~II]$ or $[O~III]$   &5008      &$[O~II]+[Fe~II]$&4957  \\
$[Fe~II]$      &5045                &$[Fe~II]$    &5008     \\
$S~II$         &5453            &$Fe~II$        &5020   \\
$Fe~II$        &5582            &$Fe~II$        &5170  	\\
$[Fe~II]$      &5655             &$Fe~II$	   &5236      	 \\
$[Fe~II]$      &5871           &$Fe~II$	   &5277      	 \\
$He~I$         &5876             &$Fe~II$	   &5318      \\
$Fe~II$        &6249            &$Fe~II$     &5536       \\
$H\alpha$   &6564             &$Ne~I$ ?	  &6165        \\
$[N~II]$       &6584          &$Fe~II$    &6249     	\\
$He~I$         &6678          &$[O~I]$    &6300       	\\
$[S~II]$       &6717           &$Fe~II$    &6458        \\
$[S~II]$       &6731        &$Fe~II$    &6518        \\
$[Ar~III]$     &7135          &$H\alpha$ &6564      \\
$He~I$         &7283             &$O~I$ ? &7774	      	 \\
$O~II$         &7321         &$O~I$	 &8446        	 \\
--           &7723           &$H~I~P17$  &8467	       \\
$S~II$         &7852          &  $Ca~II$    &8498     \\
--           &8565           &  $Ca~II$    &8542     \\
$H~I~P14$       &8598        &  $H~I~P14$  &8600    	\\
$Ca~II$        &8662 	       &  $Ca~II$    &8662      \\
$Fe~II$        &8674         &  $H~I~P12$   &8750   	\\
$Cr~II$        &8832          &  $Cr~II$    &8830    	\\
$[S~III]$ ?      &9071        &  $H~I~P11$   &8863   	\\
$Fe~II$        &9073          &  $Fe~II $    &8929   	\\
             &	              &  $H~I~P10$   &9014   	\\
  	         &                &  $H~I~P9$    &9229   	\\
\hline
\end{tabular}
\tablefoot{Controversial line identifications are marked by a question mark.}
\end{table}

\section{Discussion}

Since the main goal of this paper was to identify as many dusty massive stars as possible, the critical question is the efficiency of the selection criteria. The discovery of new RSGs and emission line stars provides an opportunity to test the success rate of our criteria. In Table \ref{tab:stat} we present the classification of targets in each program galaxy. We provide the following seven categories: (i) "Unclassified" -- targets mainly with low S/N spectra, for which we were not able to provide a classification; (ii) "Spectral type only" -- targets for which only the spectral type identification was performed; (iii) "Giants" -- foreground giants; (iv) "RSGs"; (v) "Em. line objects" -- emission line objects; (vi) "Background objects" -- quasar (WLM 26) and background galaxy (WLM 15); (vii) "Carbon stars". The number of targets satisfying our selection criteria are marked in bold font, while the number of targets that were randomly selected to fill the free space on the MXU masks are listed in parentheses. The percentage of the selected targets belonging to each of these categories is listed in the last line of the table. The percentage of evolved massive targets is 30$\%$ for RSGs and 5$\%$ for emission line objects. The accuracy of the mid-IR photometry that was used for the selection process is likely responsible for the low percentages. Also, we expect that evolved massive stars exist among the "unclassified" and "spectral type only" targets, which we could not identify due to absence of the Ca II region in the observed spectra. Moreover, among the "Unclassified" spectra there are some targets that lie on the red part of the CMDs ($[3.6]-[4.5] > 0.2$ mag). These targets could be extreme (dusty) AGB stars, especially the targets for which the spectra are very noisy.

Among our sample of dIrr galaxies the most numerous class of spectroscopically confirmed dusty massive stars is RSGs. We can compare the success of the mid-IR photometric selection criteria with the "classical" optical selection criteria \citep{M98,LM2012}. \cite{M98} presented the first attempt to provide a tool to distinguish RSG candidates from foreground stars. Using the \cite{Kurucz1992} ATLAS9 stellar atmospheres, it was shown that it is possible to separate RSGs from foreground giants from a two-color diagram ($B-V$ vs. $V-R$). The least-squares model fit, in the formalism of the $B$, $V$ and $R$ bands, for separating RSGs from foreground contaminants is: $B-V=-1.599 \times (V-R)^{2} + 4.18 \times (V-R) - 1.04 + \delta$. The empirical coefficient $\delta$ is added to fit the positions of spectroscopically confirmed RSGs in the program galaxies. \cite{M98}, \cite{M09}, and \cite{Drout12} adopted $\delta$ = 0.1 by fitting the positions of spectroscopically confirmed RSGs in NGC 6822, M31 and M33. All targets that lie above this curve are considered to be candidate RSGs, while targets below the line are considered to be foreground contaminants. The generalization of this dividing curve as a line $B-V=1.25 \times (V-R) + 0.45$ has also been used in \cite{LM2012}.

\begin{figure*}
\begin{center}
\includegraphics[width=1\linewidth]{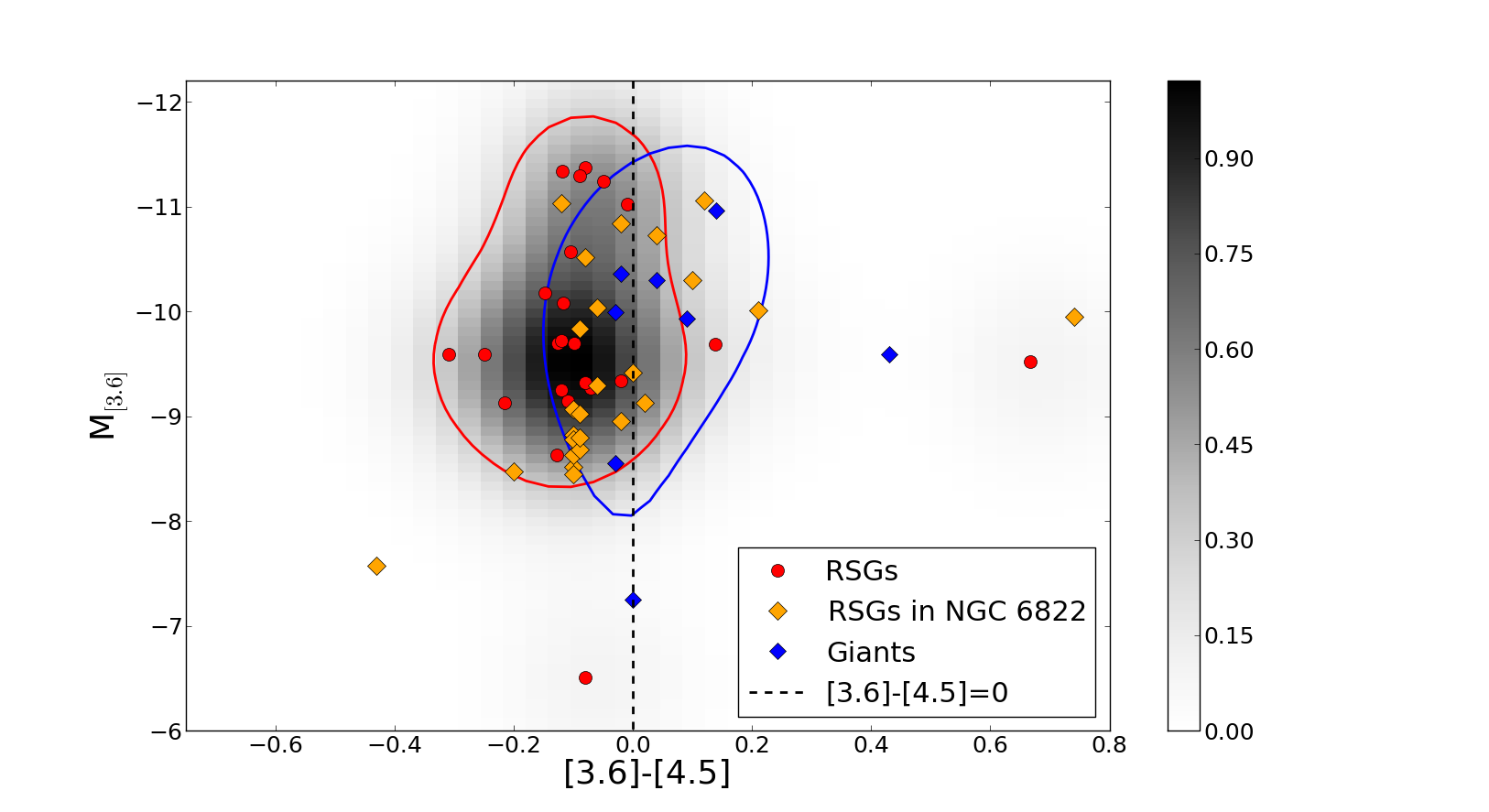}
\includegraphics[width=1\linewidth]{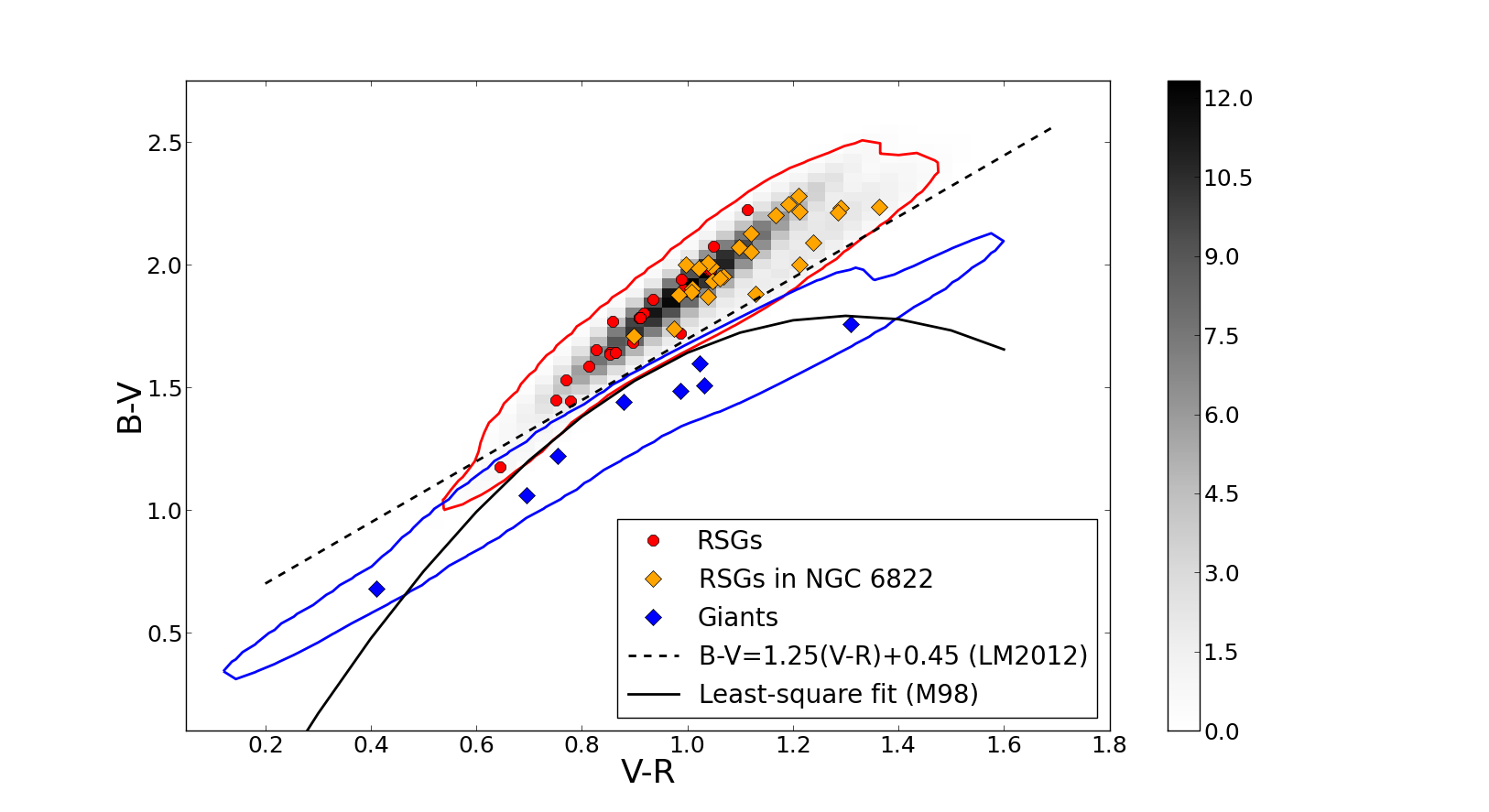}
\end{center}
\caption[]{$M_{[3.6]}$ vs. $[3.6]-[4.5]$ CMD (upper panel) and two-color optical diagram (lower panel) for all spectroscopically confirmed RSGs and giants in our 4 program galaxies, including known RSGs from literature in WLM, NGC 6822, IC 1613 and Sextans A. RSGs in NGC 6822 are marked separately, due to the different source mid-IR photometry. RSGs are labeled by red filled circles, giants are labeled by blue filled diamonds, the color contours correspond to a 1$\sigma$ dispersion of the gaussian PDF. The probability density function for the whole sample of RSGs is shown in gray, the colorbar corresponds to relative value of the PDF. The dashed line at $[3.6]-[4.5]=0$ demarcates our mid-IR selection criteria for identifying RSGs. The dashed line on the two-color diagram corresponds to the empirical dividing line that separates RSGs from foreground giant candidates defined in \cite{LM2012}. The solid line corresponds to a least-square fit model of separating RSG candidates from foreground candidates, provided in \cite{M98}.}
\label{Fig10}
\end{figure*}

In Figure \ref{Fig10} we plot all identified RSGs and giants for all program galaxies, including 2 RSGs in IC 1613, 3 in Sextans A from \cite{britavskiy14}, 11 RSGs in WLM from \cite{LM2012}, and 26 in NGC 6822 from \cite{LM2012,Patrick15}, on both the optical $B-V$ vs. $V-R$ two-color diagram and the mid-IR color-magnitude diagram. For targets in NGC 6822 we used {\it Spitzer} photometry at $[3.6]$ and $[4.5]$ bands from \cite{Spitzer_NGC6822}. Optical colors are taken from \cite{Massey_Images}. On the two-color diagram we show both the empirical parabolic curve and the line, discussed above, for separating RSGs from the giants. To compare how the positions of our RSG sample match these cut-off criteria, we calculate the probability density function (PDF) of the position of RSGs on the $B-V$ vs. $V-R$ color magnitude by using a gaussian kernel density estimation. In Figure \ref{Fig10} we indicate the contours of the 1$\sigma$ PDF of the location of RSGs and foreground giants on the mid-IR CMD (upper panel), and on the optical two-color diagram (lower panel). The colorbar corresponds to the relative value of the probability density that RSGs are located in this specific region. We find that all RSGs are grouped in the narrow region predicted on the two-color diagram, above the empirical cut-off lines. Moreover, the 1$\sigma$ interval of the PDF for the RSG region completely agrees with this parabolic equation. Our analysis indicates that both linear and parabolic cut-off lines \citep[from][]{M98} can be applied as a tool to separate RSGs from foreground giants. Taking into account that we plotted RSGs from different galaxies with slightly different metallicities, we provide evidence that this dividing line is universal and might be used for all galaxies for which precise optical photometry does exist.

\begin{figure*}
\begin{center}
\includegraphics[width=1.1\linewidth]{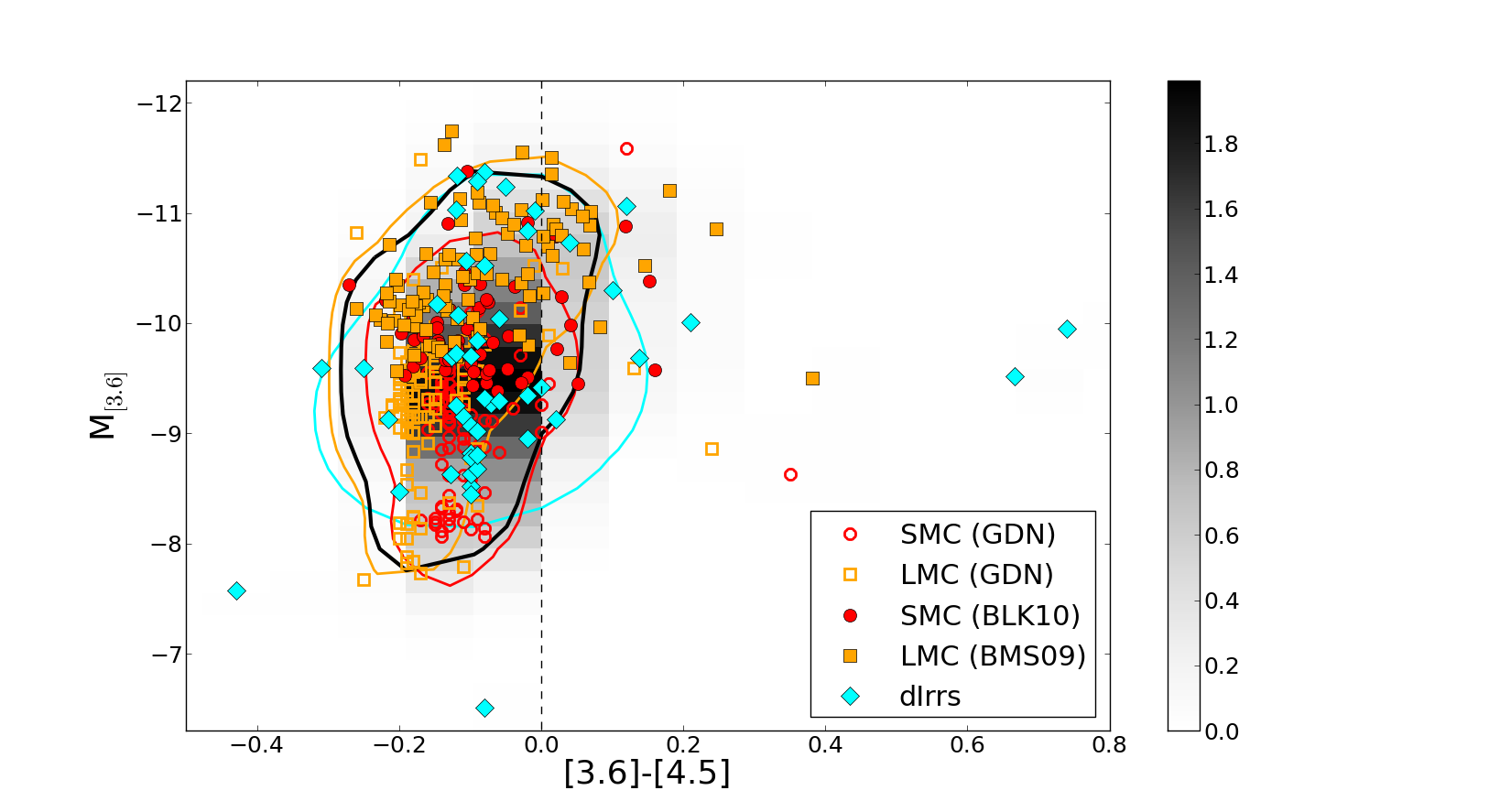}
\includegraphics[width=1.1\linewidth]{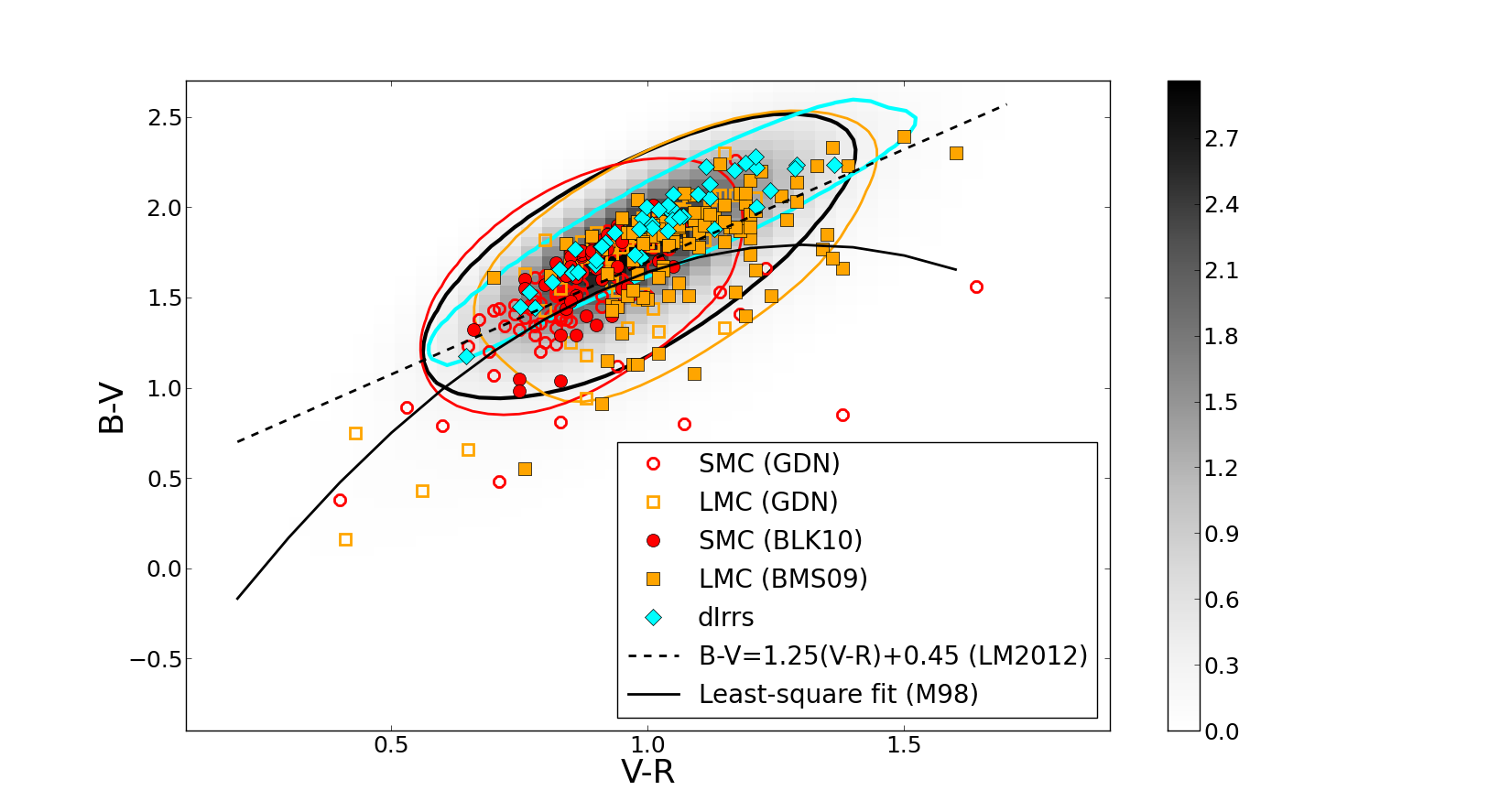}
\end{center}
\caption[]{$M_{[3.6]}$ vs. $[3.6]-[4.5]$ CMD (upper panel) and $B-V$ vs. $V-R$ diagram (lower panel) for spectroscopically confirmed RSGs in 6 Local Group dIrrs (Phoenix, Pegasus, Sextans A, WLM, IC 1613 and NGC 6822), the LMC, and the SMC. The RSGs from \cite{new_smclmc} have a label "GDN"; RSGs with labels "BLK10" and "BMS09" are from \cite{BLK10,BMS09} respectively.
The color contours correspond to a 1$\sigma$ dispersion of the PDF for all RSGs for each galaxy, respectively. The black contour corresponds to a 1$\sigma$ dispersion for the PDF of all RSGs in listed galaxies. The probability density function for all RSGs is marked in gray, the colorbar corresponds to relative values of the PDF. The dashed line at $[3.6]-[4.5]=0$ demarcates our selection criteria for identifying RSGs. The dashed line and the solid line on the two-color diagram have the same meaning as in Figure \ref{Fig10}. Optical photometry for RSGs in the Magellanic Clouds is from \cite{Massey_2002_MC}.}
\label{Fig11}
\end{figure*}

The situation is different with the mid-IR selection criteria. As we see from the upper panel of Figure \ref{Fig10}, the RSGs have a large dispersion, and a high probability of foreground contaminants (as shown by the 1$\sigma$ contour of the PDF of the foreground giants from our sample). These effects make our mid-IR selection criteria less efficient than the optical colors as a tool for separating RSGs from foreground giants. Furthermore, the cutoff line $[3.6]-[4.5]<0$ that we used for selection, does not include all spectroscopically confirmed RSGs. To generalize this statement we plotted all spectroscopically confirmed RSGs in the SMC: 59 from \cite{BLK10}, 83 from \cite{new_smclmc}, and the LMC: 96 from \cite{BMS09}, 96 from \cite{new_smclmc} together with RSGs from our sample of dIrr galaxies in Figure \ref{Fig11}. We plot the WISE W1 (3.353 $\mu m$) and W2 (4.603 $\mu m$) photometry of the RSGs from \cite{new_smclmc}, which are very close to the {\it Spitzer} [3.6] and [4.5] bands.
We find the majority of new RSGs in the SMC and the LMC to be fainter compared with the previously known RSGs. The difference in absolute magnitude is likely due to the fact that the WISE photometry has better sensitivity, despite worse resolution than the {\it Spitzer} photometry for the Magellanic Clouds. Also, there is a difference in the $[3.6]-[4.5]$ colors for the RSGs in the LMC vs. SMC. The maximum (peak) of the PDF for RSGs from the LMC has a bluer color ($[3.6]-[4.5] = -0.17$ mag) and brighter magnitude (M$_{[3.6]} = -10.15$ mag) than the position of the maximum PDF for the RSGs from the SMC ($[3.6]-[4.5] = -0.02$ mag, M$_{[3.6]} = -9.51$ mag). This difference, originates in the different metallicities of the LMC and the SMC, which affect the depth of CO bands at [4.5] $\mu m$, and as a result the $[3.6]-[4.5]$ color.

We added the same sample of RSGs from the Magellanic Clouds on the optical two-color diagram (lower panel of Figure \ref{Fig11}). Optical colors for this sample of RSGs are taken from \cite{Massey_2002_MC}. For such a big sample of RSGs we can see a big scatter of their position on the two-color diagram, which is mainly due to the accuracy of the optical photometry. Nevertheless, the majority ($\approx$ 70 $\%$) of spectroscopically confirmed RSGs satisfy the optical selection criteria. Comparing this success rate in the optical selection criteria with a number of RSGs that satisfy the mid-IR selection criteria ($\approx$ 90 $\%$), we can conclude that the mid-IR selection criteria are useful for selecting dusty massive stars, however these criteria are not so efficient for separating them from foreground contaminants, in contrast with the optical selection criteria (as shown in Figure \ref{Fig10}).

We conclude that RSGs group in a broader region in the mid-IR CMD, in contrast to the narrow region defined on the optical two-color diagram, which makes optical selection criteria more reliable, when high-quality optical photometry exists for the investigated galaxies. However, the mid-IR criteria are useful for the large number of galaxies with {\it Spitzer} imaging that lack deep optical photometry. This work together with \cite{britavskiy14}, increased the sample of spectroscopically confirmed RSGs in dIrr galaxies in the Local Group by 12 (27 $\%$): we have identified 2 RSGs in IC 1613, 7 RSGs in Sextans A, 1 RSG in Phoenix, 2 RSGs in Pegasus. Prior to these works, there where 44 RSGs spectroscopically confirmed in dIrrs of the Local Group: 33 RSGs were known in NGC 6822 \citep{M98,LM2012,Patrick15} and 11 RSGs were known in WLM \citep{LM2012}.



\begin{table*}
\caption{Summary classification of our 79 observed targets.}
\label{tab:stat}
\begin{tabular}{@{}lllllllll@{}}
\hline\hline
ID    & All observed & Unclassified & Spectral type    & Giants &   RSGs & Em. line  & Background        & Carbon stars\\
      & targets      &              &  only      &        &        & objects    & objects         &             \\
\hline
Pegasus & {\bf 11} (+8)    &    {\bf 2} (+7)    &  {\bf 3}       &     {\bf 4}       &     {\bf 2}       &      {\bf 0}     & {\bf 0} & (+1)\\
Phoenix &  {\bf 2} (+12)   &    {\bf 0} (+5)    &  {\bf 2} (+2)  &     {\bf 0} (+2)  &     {\bf 0} (+1)  &      {\bf 0}     & {\bf 0} & (+2)\\
Sextans A & {\bf 15}       &    {\bf 5}   &  {\bf 2}       &     {\bf 1}    &     {\bf 7}       &      {\bf 0}     &{\bf  0} & (0)  \\
WLM   & {\bf 15} (+16)     &   {\bf 5} (+8)    &  {\bf 3} (+1)   &     {\bf 0} (+1)       &     {\bf 4}  &      {\bf 2} & {\bf 1} (+1)& (+5)  \\
\hline									        	    					
Total     &   {\bf 43} (+36)   &    {\bf 12} (+20) & {\bf 10} (+3) &  {\bf 5} (+3)  &  {\bf 13} (+1)  &   {\bf  2}  & {\bf 1} (+1) & (+8)\\
 $\%$      &  {\bf 100}  &   {\bf  28}  &  {\bf 23 }       &  {\bf 12}   &    {\bf 30}   &   {\bf  5}   &  {\bf 2}& --  \\
\hline
\end{tabular}
\tablefoot{
The number of targets that satisfy our selection criteria are marked by bold font. Targets that were randomly selected to fill the MXU slits are given in parentheses.}
\end{table*}

\section{Summary}

In this work we present an attempt to complete the census of dusty massive stars in a sample of four dwarf irregular galaxies in the Local Group, namely Pegasus, Phoenix, Sextans A, and WLM. We used mid-IR selection criteria based on the $[3.6]$ and $[4.5]$ bands for identifying dusty types of massive stars such as RSGs, LBVs, and sgB[e] stars. We performed a spectroscopic analysis for 79 sources. As a result of our work 13 RSGs were identified (6 of them are newly discovered), 2 emission line objects (one of them is a newly identified iron star), and 1 candidate yellow supergiant. The large sample of spectroscopically defined RSGs gave us the possibility to investigate and revise the mid-IR and optical selection criteria. We provide evidence that the most precise method to select RSGs from foreground objects is an optical $B-V$ vs. $V-R$ color-color diagram with the linear and the parabolic separation line of \cite{M98}. The mid-IR selection criteria have lower efficiency, however, with available {\it Spitzer} mid-IR photometry \citep{dustings} for 13 dwarf irregular galaxies in the Local Group, our tool provides an opportunity to select candidate dusty massive stars in star-forming dIrrs in cases where accurate optical photometry does not exist or is not deep enough. The mid-IR selection criteria can be applied both to galaxies in the Local Group and to more distant galaxies \citep[e.g. M83;][]{Stephen15} by using mid-infrared photometry from other surveys, e.g. the Local Volume Legacy \citep{LVL} and the SIRTF Nearby Galaxy Survey \citep{SIRTF}. At this stage, the verification of selection criteria of emission line objects is not possible, since only a small number of such stars were identified. This work provides a basis for future identification of dusty massive stars in the Local Group. Moreover, newly discovered RSGs can be used for investigating metallicities and the properties of their host galaxies \citep[e.g.][]{Davies_2013,Patrick15}.

\section*{Acknowledgments}
We thank the anonymous referee for helpful comments that have improved the manuscript. N. Britavskiy and A.Z. Bonanos acknowledge funding by the European Union (European Social Fund) and National Resources under the "ARISTEIA" action of the Operational Programme "Education and Lifelong Learning" in Greece.
We would like to thank A. Miroshnichenko for useful discussions on the identification of emission lines in spectra.
This research has made use of NASA's Astrophysics Data System Bibliographic Services and the VizieR catalogue access tool, CDS, Strasbourg, France. Funding for SDSS-III has been provided by the Alfred P. Sloan Foundation, the Participating Institutions, the National Science Foundation, and the U.S. Department of Energy Office of Science.


\end{document}